\documentclass{nature}
\usepackage{graphicx}
\makeatletter
\let\saved@includegraphics\includegraphics
\AtBeginDocument{\let\includegraphics\saved@includegraphics}
\renewenvironment*{figure}{\@float{figure}}{\end@float}
\makeatother
\usepackage{epstopdf}
\usepackage{dcolumn}
\usepackage{bm}
\usepackage{amssymb}
\usepackage{color}
\usepackage{amsmath}

\usepackage{soul}
\usepackage{float}

\usepackage{cancel}
\newcommand{\beq}{\begin{eqnarray}}
\newcommand{\eeq}{\end{eqnarray}}

\title{Locally and globally chiral fields for ultimate control of chiral light matter interaction}

\author{David Ayuso$^{1}$,
Ofer Neufeld$^{2}$,
Andres F. Ordonez$^{1,3}$,
Piero Decleva$^{4}$,
Gavriel Lerner$^{2}$,
Oren Cohen$^{2}$,
Misha Ivanov$^{1,5,6}$,
Olga Smirnova$^{1,3}$
} 

\begin{document}
\maketitle
\begin{affiliations}
\item Max-Born-Institut, Max-Born-Str. 2A, 12489 Berlin, Germany
\item Physics Department and Solid State Institute, Technion - Israel Institute of Technology, Haifa 32000, Israel
\item Technische Universit\"at Berlin, Ernst-Ruska-Geb\"aude, Hardenbergstra{\ss}e 36A, 10623 Berlin, Germany
\item Dipartimento di Scienze Chimiche e Farmaceutiche, Universit\`a degli Studi di Trieste, via L. Giorgieri 1, 34127 Trieste, Italy
\item Institute f\"ur Physik, Humboldt-Universität zu Berlin, Newtonstra{\ss}e 15, D-12489 Berlin, German
\item Department of Physics, Imperial College London, South Kensington Campus, SW72AZ London, UK
\end{affiliations}

\begin{abstract}
Light is one of the most powerful and precise tools allowing us to control\cite{Brumer_Shapiro,Karczmarek1999},
shape\cite{Matthews2018,Basov2017} and create new phases\cite{Khemani2016PRL,Zhang2017} of matter.
In this task, the magnetic component of a light wave has so far played a unique role in defining the wave's helicity, but its influence on the optical response of matter is weak. 
Chiral molecules offer a typical example where the weakness of magnetic interactions hampers our ability to control the strength of their chiral optical response\cite{book_ComprehensiveChiropticalSpectroscopy}.
It is limited several orders of magnitude below the full potential.
Here we introduce freely propagating locally and globally chiral electric fields, which interact  with chiral quantum systems extremely efficiently.
To demonstrate the degree of control enabled by such fields, we focus on the  nonlinear optical response of randomly oriented chiral molecules.
We show full control over intensity, polarization and propagation direction of the chiral optical response, enabling its background-free detection.
This response can be  fully suppressed or enhanced at will depending on the molecular handedness, achieving the ultimate limit in chiral discrimination. 
Our findings open a way to extremely efficient control of chiral  matter and to ultrafast imaging of chiral structure and dynamics in gases, liquids and solids. 
\end{abstract}



Chirality is a ubiquitous property of matter, from its elementary constituents to molecules, solids, and biological species.
Chiral molecules appear in pairs of left and right handed enantiomers, where their nuclei arrangements present two non-superimposable mirror twins.
The handedness of some macroscopic objects can be identified using a mirror, but how does one  distinguish the handedness of a microscopic chiral object?

One powerful strategy that uses light relies on the concept of ``chiral observer'' \cite{Ordonez2018_PRA}. This concept unites several revolutionary techniques of chiral discrimination \cite{Ritchie1976, Powis2000JCP,Bowering2001PRL,Fischer2005, Patterson2013Nature,Beaulieu2018NatPhys,Lux2012Angewandte, Lehmann2013JCP,yachmenev2016detecting,Comby2016JPCL,Beaulieu2016NJP,Kastner2016,Goetz2017JCP,Tutunnikov_2018}, which analyze experimental observables with respect to a chiral reference frame defined by the experimental setup.
Two axes of the reference frame are fixed by two non-collinear electric field vectors interrogating the chiral system, such as the two components of a circularly polarized laser pulse. The third laboratory axis is defined by the direction of observation, for example, by the position of the photo-electron detectors in photo-electron circular dichroism measurements\cite{Ritchie1976, Powis2000JCP,Bowering2001PRL,Beaulieu2018NatPhys,Lux2012Angewandte, Lehmann2013JCP,nahon2015valence,Comby2016JPCL,Beaulieu2016NJP,Goetz2017JCP}.

Probing chirality can also involve interaction between two chiral objects, one of them with known handedness. This  principle relies on using a well-characterized “chiral reagent” -- another chiral molecule, or chiral light.

There is a fundamental difference between a chiral observer and a chiral reagent \cite{Ordonez2018_PRA}. 
A chiral observer detects opposite directions or  phases \cite{Patterson2013Nature,Beaulieu2018NatPhys} of signals excited in the two enantiomers,
such as the opposite photo-electron currents generated by photo-ionization with circularly polarized light \cite{nahon2015valence}.
But it does not influence the total signal, i.e. the total number of photo-electrons.
In contrast, a chiral reagent leads to different total signal intensity in left and right enantiomers, such as in standard absorption circular dichroism in chiral media.
Gaining control over intensity of light emission or absorption by chiral matter requires a chiral photonic reagent -- chiral light.

Circularly polarized light, while chiral, is ill-suited for this purpose.  
The pitch of the light helix in the optical domain is too large compared to the size of the molecule, hindering chiro-optical response.  
To control chiral objects much smaller than the light wavelength \emph{efficiently}, one needs a field that is chiral \emph{locally}, in the electric-dipole approximation, 
without invoking light evolution in space.

Here we introduce, characterize and use freely propagating \emph{locally} chiral electromagnetic fields, which also maintain their handedness \emph{globally} in space.
Their global chirality map can be engineered by tuning their handedness locally, at every point.
We show that such fields enable the highest possible degree of control over the chiral non-linear optical response, quenching it in one enantiomer, while maximizing it in its mirror twin.

In locally chiral fields, the tip of the electric field vector  follows a chiral trajectory  in time (Fig. 1a). Here time plays a role of an additional dimension and a locally chiral field can be viewed as a "geometrical" object local in space, but non-local in time. Any symmetry analysis applicable to chiral molecules can be applied to locally chiral light fields, once dynamical symmetries describing its evolution in time are included  \cite{neufeld_symmetries_2017}. 
However, a symmetry analysis does not allow us to quantify their handedness.
A previously employed measure of optical chirality\cite{tang_prl_2010} is equivalent to light helicity and vanishes in the electric-dipole approximation. To quantify the handedness of locally chiral fields we need a new approach.
We take snapshots of the electric field vector at three different instants of time and evaluate the triple product of these three vectors, forming a ubiquitous chiral measure \cite{Fischer2005,Eibenberger2017PRL,Beaulieu2018NatPhys,Ordonez2018_PRA}. 
Once this triple product is averaged over time to sample the entire trajectory, we find that this measure presents a special, chiral, three-point correlation function (see Methods), which distinguishes locally chiral fields.
Thanks to the triple product, it forms a pseudoscalar, which changes sign if we mirror-reflect the field trajectory, distinguishing left-handed and right-handed locally chiral fields. What's more, it describes the interaction of locally chiral fields with matter. 

The lowest order chiral field-correlation function in the frequency domain is given by 
\begin{eqnarray}\label{H3}
h^{(3)}(-\omega_1-\omega_2,\omega_1,\omega_2) \equiv  \mathbf{F}^{*}(\omega_1+\omega_2) \cdot  \left[ \mathbf{F}(\omega_1) \times \mathbf{F}(\omega_2)\right] .
\end{eqnarray}
and its complex conjugate counterpart (see Methods), where $\mathbf{F}(\omega_i)$ are the Fourier components of the electric field vector at three different frequencies.
A non-zero triple product means that the field trajectory defines a local chiral reference frame via the three non-coplanar vectors $\mathbf{F}(\omega_1)$, $\mathbf{F}(\omega_2)$ and $\mathbf{F}(\omega_1+\omega_2)$.
This is why to have non-zero $h^{(3)}$, a locally chiral field needs at least three different frequency components.
$h^{(3)}$ governs perturbative light-matter interaction, such as absorption circular dichroism induced solely by laser electric fields (see Methods) and enantiosensitive population of rotational states in chiral molecules\cite{Eibenberger2017PRL} using microwave fields.

\begin{figure}
\centering
\includegraphics[width=\linewidth, keepaspectratio=true]{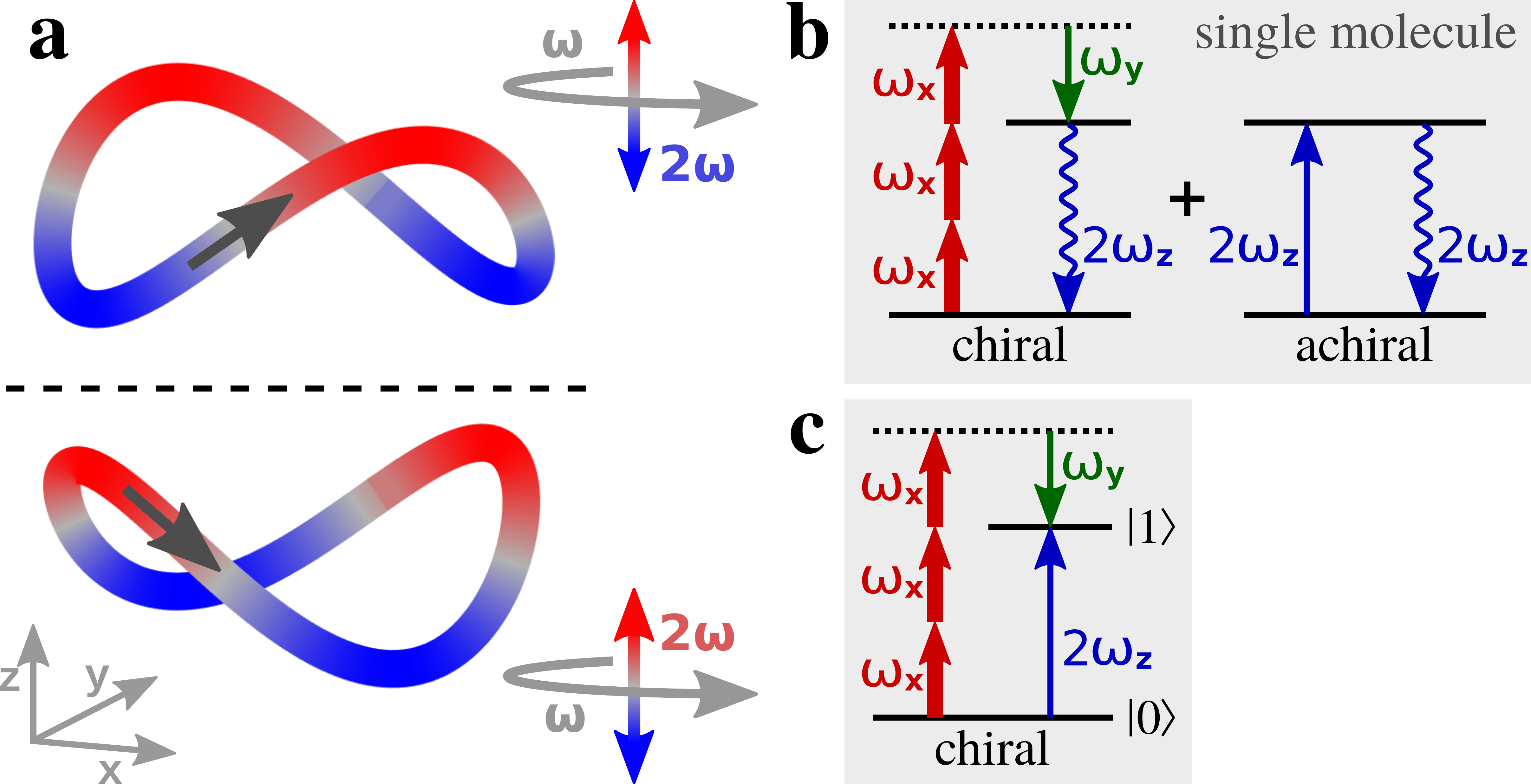}
\caption{Control with locally chiral fields.
\textbf{a,} Trajectory of the locally chiral field in Eq. \eqref{eq_field}, color shows $z$ coordinate.
Reflection through the $xy$ plane changes the sign of the $z$ field component and thus the field's handedness. Inset shows that the field is a superposition of a component of frequency $\omega$ elliptically polarized in the $xy$ plane, and a component of frequency $2\omega$ linearly polarized along $z$ (see Eq. \eqref{eq_field}).
\textbf{b,} Interference of chiral and achiral pathways in even (second) harmonic generation from a single molecule controlled by $h^{(5)}$.
\textbf{c,} Interfering multiphoton pathways controlled by $h^{(5)}$ resulting in enantiosensitive absorption into level $|1\rangle$.
}
\label{fig_scheme}
\end{figure}

One may think that if $h^{(3)}=0$, then the field is not locally chiral. This is not true, $h^{(3)}\neq 0$ is
a sufficient but not necessary condition to have a locally chiral field.
One can easily verify that the electric field vector of the very simple two-color field (Fig. \ref{fig_scheme}a):
\begin{equation}\label{eq_field}
F_{\omega} \cos{(\omega t)}\mathbf{x} + \varepsilon F_{\omega} \sin{(\omega t)}\mathbf{y} + F_{2\omega} \cos{(2\omega t + \phi_{\omega,2\omega})}\mathbf{z}
\end{equation}
traces a chiral trajectory in time, even though its $h^{(3)}=0$.
Its chirality  manifests in higher order time-correlations, which can be perfectly put to use in higher order non-linear light matter interactions.
Indeed, the next order correlation function for this field (see Methods)
\begin{equation}\label{h5_even}
h^{(5)}(-2\omega,-\omega,\omega, \omega,\omega)\equiv \left\{\mathbf{F}^{*}(2\omega) \cdot [\mathbf{F}^{*}(\omega)\times\mathbf{F}(\omega)] \right\}  \left[\mathbf{F}(\omega)\cdot\mathbf{F}(\omega)\right]
\end{equation}
is non-zero.
Eq. (\ref{h5_even}) shows that the $\omega$-$2\omega$ field also forms a local chiral coordinate frame, and  we can identify three non-coplanar vectors upon temporal averaging.
Here, the system absorbs three photons of frequency $\omega$ with $x$ polarization, emits one photon of frequency $\omega$ with $y$ polarization and emits one $2\omega$ photon with $z$ polarization.
The handedness $h^{(5)}$ of this locally chiral field depends on $\phi_{\omega,2\omega}$, the $\omega$-$2\omega$ phase delay.
The spatial map of such delays is under our full control.
It determines the $h^{(5)}$ map globally in space.
Thus, we now have a chiral photonic reagent, whose handedness can be tailored locally to address specific regions in space and specific multiphoton orders of interactions, with extremely high efficiency.

The mechanism of control over enantiosensitive optical response in the perturbative multiphoton regime is as follows.
In isotropic chiral media, in the electric-dipole approximation the chiral response occurs only due to interactions involving an even number of photons both the in weak-field\cite{Fischer2005} and strong-field\cite{Neufeld2018_arxiv} regimes, but the non-chiral contribution must involve an odd number of photons, just like in ordinary isotropic media. Note that this difference allows one to differentiate between chiral and non-chiral media in the intensity of non-linear optical response associated with even number of absorbed photons\cite{Neufeld2018_arxiv}. However, this intensity is the same for right-handed and left-handed enantiomers, and their response differs only by a global phase \cite{Neufeld2018_arxiv, Patterson2013Nature}. In contrast, locally chiral fields make the signal intensity enantio-sensitive and allow one to control its strength depending on the molecular handedness. 

Indeed, consider the intensity of the optical response triggered by the $\omega$-$2\omega$ field (Fig. \ref{fig_scheme}a) at frequency $2\omega$. Since the enantiosensitive second order susceptibility $\chi^{(2)}(2\omega;\omega,\omega)=0$ (see e.g.\cite{Fischer2005}), the lowest order enantio-sensitive multiphoton process involves four photons of frequency $\omega$ in the chiral channel and one photon of frequency $2\omega$ in the achiral channel:
\begin{eqnarray}\label{interf}
\left\vert\mathbf{P}(2\omega)\right\vert^{2} &=& \left\vert \chi^{(4)} \left[\mathbf{F}^{*}(\omega) \times \mathbf{F}(\omega)\right]\left[\mathbf{F}(\omega)\cdot\mathbf{F}(\omega)\right] + \chi^{(1)}\mathbf{F}(2\omega)\right\vert^2  \nonumber \\
 &=& (\mathrm{diagonal\,\,terms}) + \chi^{(1)*}\chi^{(4)}h^{(5)} + \mathrm{c.c.}
\end{eqnarray}
Here $\mathbf{P}(2\omega)$ is the induced polarization at $2\omega$, $\chi^{(4)}=\sigma_{\mathrm{M}}\vert \chi^{(4)} \vert e^{i\phi_{4}}$ is the enantiosensitive fourth-order susceptibility (see Methods), and $\chi^{(1)}=\vert \chi^{(1)} \vert e^{i\phi_{1}}$ is the achiral linear susceptibility.
The interference term 
\begin{eqnarray}\label{interf1}
2\sigma_{\mathrm{M}} \sigma_{\mathrm{L}} \big|\chi^{(1)}\big|\big|\chi^{(4)}\big|\big|h^{(5)}\big|\cos(\phi_{\mathrm{M}}+\phi_{\omega,2\omega})
\end{eqnarray}
is controlled by the chiral-field correlation function $h^{(5)}=\sigma_{\mathrm{L}}\vert h^{(5)}\vert e^{i{\phi_{\omega,2\omega}}}$.
It depends on the molecular phase $\phi_{\mathrm{M}}=\phi_{4}-\phi_{1}$ associated with complex susceptibilities, the relative phase $\phi_{\omega,2\omega}$ between the $\omega$ and $2\omega$ field components,
the handedness of the molecule $\sigma_{\mathrm{M}}=\pm1$, and the 
sense of in-plane rotation of light $\sigma_{\mathrm{L}}=\pm1$.
Eqs. (\ref{interf},\ref{interf1}) show that tuning the strengths and the relative phase $\phi_{\omega,2\omega}$ between the $\omega$ and $2\omega$ fields we can achieve perfect constructive/destructive interference and fully suppress or maximally enhance the signal intensity in the selected enantiomer (see Fig. \ref{fig_scheme}b).
No additional achiral background channels are allowed due to selection rules.
The exact same interference, as is clear from Fig. \ref{fig_scheme}c, controls absorption of the $2\omega$ field.

\begin{figure}
\centering
\includegraphics[width=\linewidth, keepaspectratio=true]{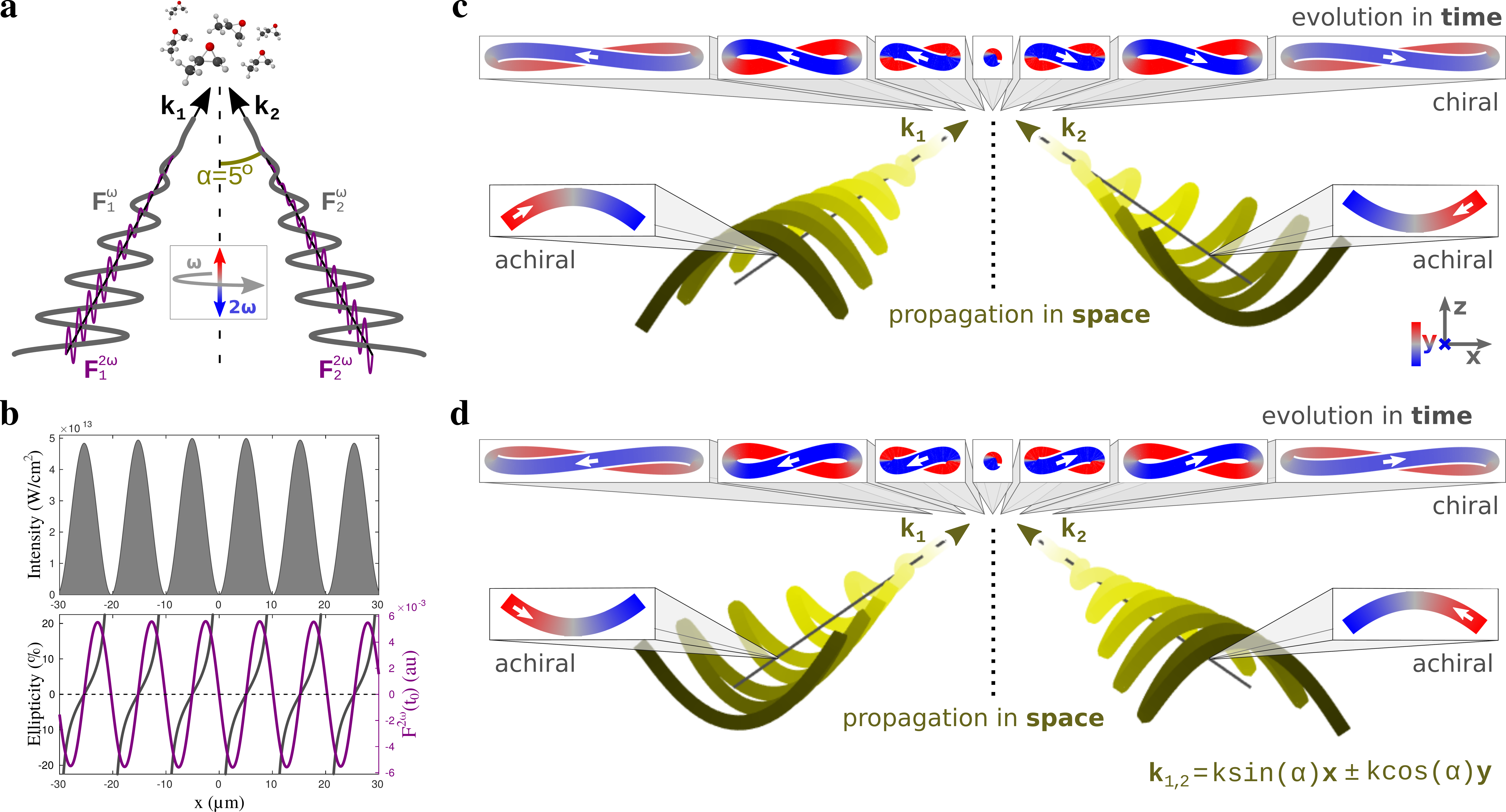}
\caption{Locally and globally chiral field.
\textbf{a}, The setup for generating a locally and globally chiral field includes two non-collinear laser beams, 
each carrying a strong $\omega$ field and a weak, orthogonally polarized, $2\omega$ field.
The total $\omega$ field is elliptical in the $xy$ plane, the $2\omega$ field is linear along $z$.
\textbf{b}, Ellipticity of $\omega$ field (grey) and amplitude of the  $2\omega$ field (purple) across the focus. 
The ellipticity flips sign at the same position in the focus where the $2\omega$ field changes its oscillation phase by $\pi$, ensuring that this locally chiral field maintains its handedness globally in space.
Panel \textbf{c} shows that the chiral temporal structure at different points across the focus, shown in top-row boxes, maintains its handedness.
\textbf{d}, same as \textbf{c} but for the opposite enantiomer of the field.}
\label{fig_global}
\end{figure}

We now turn to quantitative analysis and focus on high harmonic generation, a ubiquitous process in gases, liquids and solids.
The required locally chiral field shown in Fig. \ref{fig_scheme}a can be easily generated using the setup in Fig. \ref{fig_global}a.
It involves two non-collinear beams with wavevectors $\mathbf{k}_1$ and $\mathbf{k}_2$ propagating in the $xy$ plane at small angles $\pm \alpha$ to the $y$ axis (here $\alpha=5^{\circ}$).
Each beam is made of linearly polarized $\omega$ and $2\omega$ fields  with orthogonal polarizations and controlled phase delays (see Methods).
Thanks to the non-zero $\alpha$, the $\omega$ field is elliptically polarized in the $xy$ plane, with the minor component along the propagation direction $y$.
The handedness of the locally chiral field does not change globally in space: 
the amplitude of the $2\omega$ field, polarized along $z$, and the ellipticity of $\omega$ field, confined in the $xy$ plane, flip sign at the same positions (see Fig. \ref{fig_global}b and Supplementary Information). 
The chiral structure of the field trajectory in time maintains its handedness across the focus, as can be seen in the top rows of Fig. 2(c,d). 

\begin{figure}
\centering
\includegraphics[width=\linewidth, keepaspectratio=true]{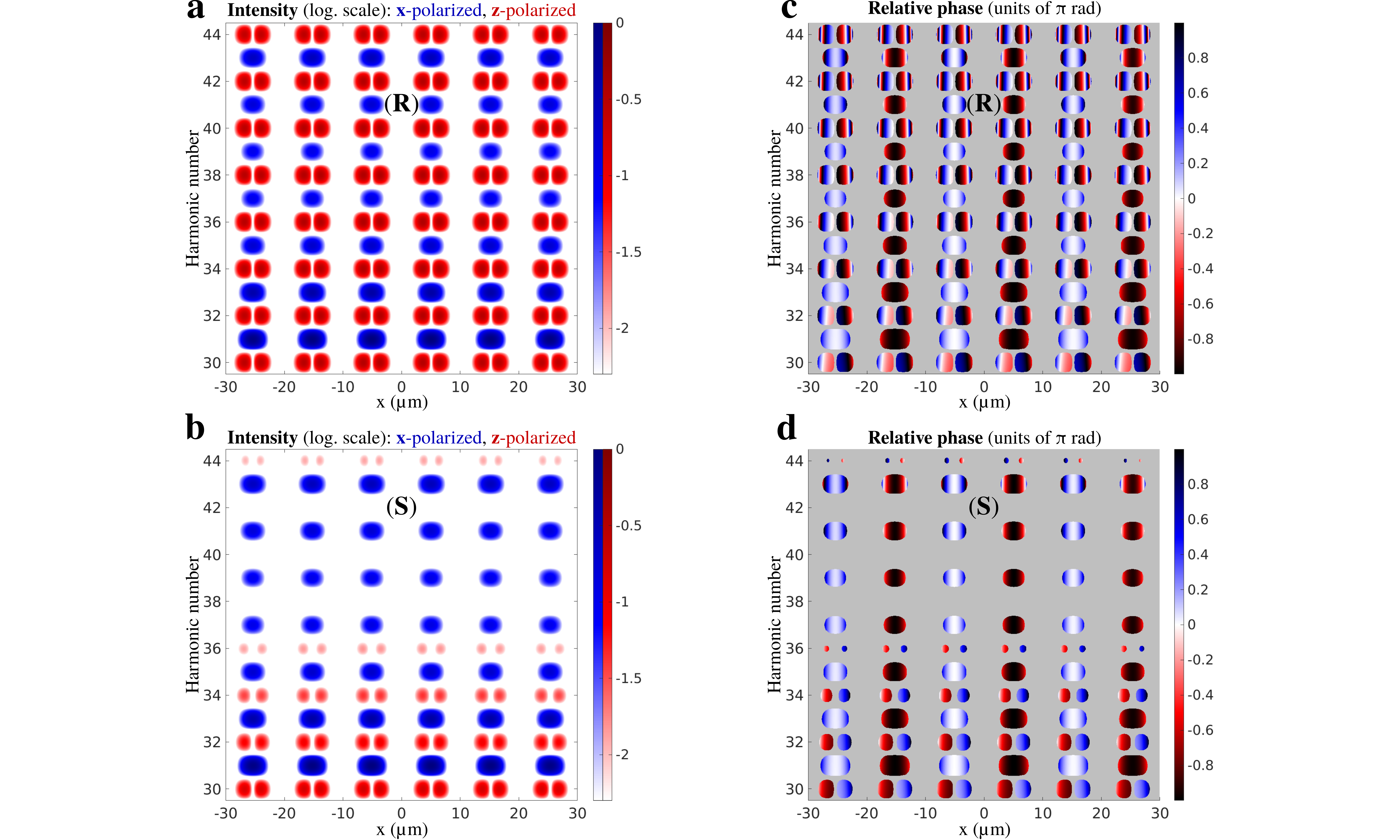}
\caption{Near field enantio-sensitive high harmonic generation by a locally chiral field.
Enantiosensitive polarization grating created in enantiopure samples of $R$ and $S$ propylene oxide molecules: intensity (a, b) and phase (c, d).
The fundamental wavelength is $\lambda=1.77\mu$m, intensity $I_\omega=1.3\cdot10^{13}$W$/$cm$^{2}$,  $I_{2\omega}=1\%I_{\omega}$, $\phi_{\omega,2\omega}=0$, pulse duration 23 fsec at constant intensity, $\alpha=5^{\circ}$, and focal diameter $400\mu$m.
}
\label{fig_near}
\end{figure}

To describe the non-linear response of an isotropic chiral medium, we developed a quantitative model of high harmonic response in randomly oriented propylene oxide, and verified it against experimental results of Ref.\cite{Cireasa2015NatPhys} (see Supplementary Fig. 1 (a-c)), with excellent agreement. 

The periodic locally chiral structure of the field along the $x$ axis leads to the amplitude (Figs. \ref{fig_near}a,b) and phase (Figs. \ref{fig_near}c,d) gratings of the generated chiral response.
Here the intensity of the second harmonic field has been set to $1\%$ of the fundamental, and  $\phi_{\omega,2\omega}=0$. The gratings in Figs. \ref{fig_near}a-d are completely different for the right and left enantiomers, demonstrating enantio-sensitive intensity
of the optical response already at the single-molecule level.

\begin{figure}
\centering
\includegraphics[width=\linewidth, keepaspectratio=true]{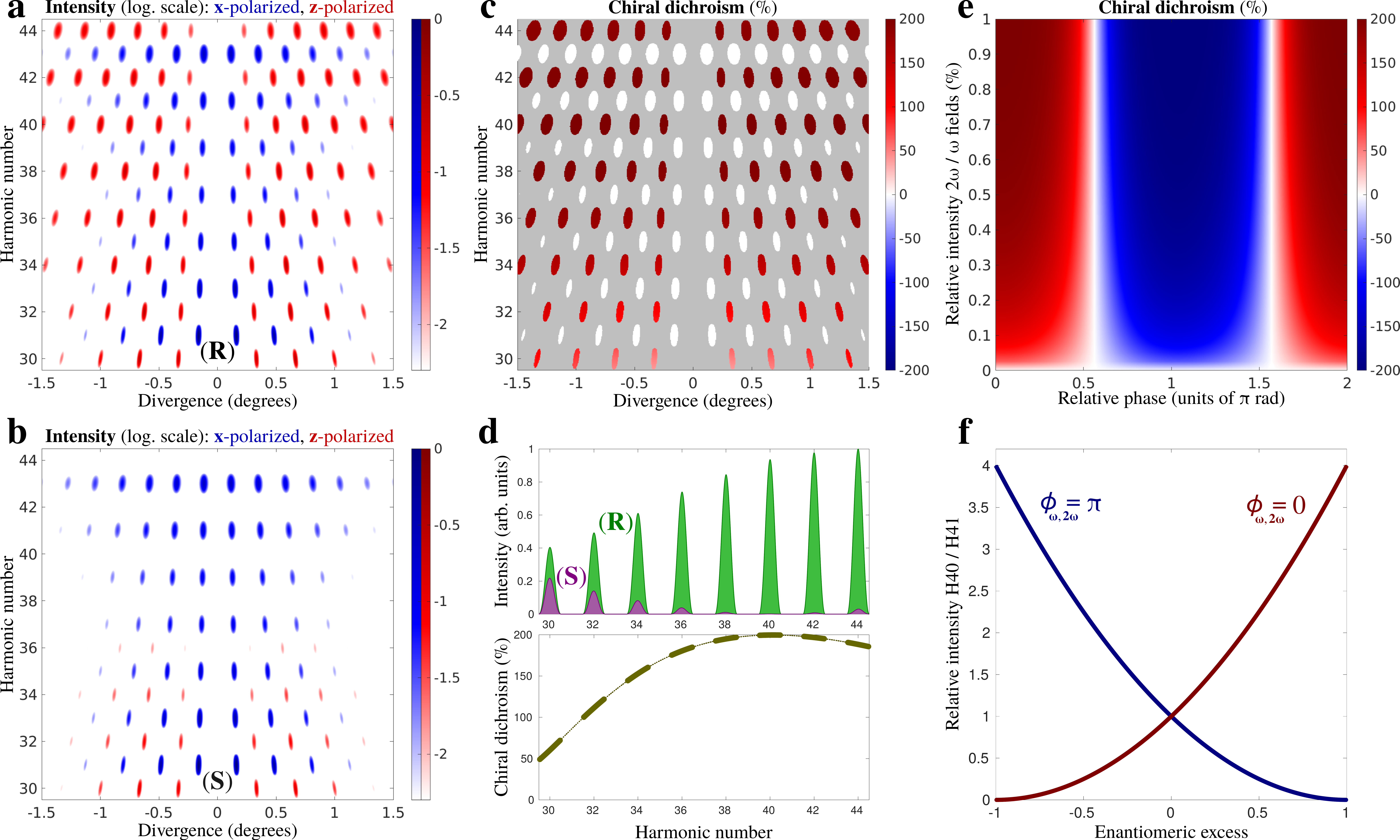}
\caption{Far-field harmonic intensity and chiral dichroism.
\textbf{a, b,} Harmonic intensity for randomly oriented enantiopure $R$ and $S$ propylene oxide molecules, for the same field as in Fig. 3, $I_{2\omega}/I_{\omega}=0.01$ and $\phi_{\omega,2\omega}=0$.
Shifting $\phi_{\omega,2\omega}$ by $\pi$ is equivalent to exchanging the enantiomer.
\textbf{c,} Chiral dichroism in harmonic intensity.
\textbf{d,} Total angle-integrated even harmonic intensity and chiral dichroism.
\textbf{e,} Chiral dichroism in H40 versus $\phi_{\omega,2\omega}$.
\textbf{f,} Intensity ratio between H40 and H41 as a function of the enantiomeric excess when $I_{2\omega}/I_{\omega}=0.01$ and $\phi_{\omega,2\omega}=0$ (red) and $\phi_{\omega,2\omega}=\pi$ (blue).
}
\label{fig_far}
\end{figure}

Figs. \ref{fig_far}a,b show how the enantiosensitive gratings translate into the far field (see Methods). Even and odd harmonics are emitted in different directions.
Their spatial separation follows from momentum conservation upon net absorption of the corresponding number of photons. 
Thus, even harmonics constitute a background-free measurement of the molecular handedness, separated from the achiral signal  in frequency, polarization and space.
Note that, in isotropic media, the harmonic signal at even frequencies of the fundamental can only be $z$-polarized, but odd harmonics are polarized along the $x$ axis.

The intensity of even harmonics is  determined by the interference between chiral and achiral pathways $\propto \cos{(\phi_M+f_{N}(\phi_{\omega,2\omega}))}$,
where the controlled phase $f_N(\phi_{\omega,2\omega})$ is a non-linear function of $\phi_{\omega,2\omega}$, for every harmonic number (see Supplementary Information).
This interference can be controlled by changing $\phi_{\omega,2\omega}$ and the 
intensity of the $2\omega$ field.
We can therefore selectively quench the harmonic emission from the $S$ enantiomer and enhance the signal of its mirror image, yielding unprecedented enantiosensitive response.
The signal $I_S$  from the left handed molecule is completely suppressed for H40,  the chiral dichroism reaches $200\%$ (see Fig. \ref{fig_far}c,d).
Shifting $\phi_{\omega,2\omega}$  by $\pi$ we reverse the handedness of the driver and thus obtain the exactly  opposite result: suppression of $I_R$ and enhancement of $I_S$ (see Fig. \ref{fig_far}e). The harmonic number(s) that exhibit $200\%$ chiral dichroism can be selected by tuning the parameters of the second harmonic field and used for concentration-independent determination of both the amplitude and the sign of the enantiomeric excess $ee$ in macroscopic mixtures from simple intensity measurements (see Fig. \ref{fig_far}f).
Indeed, for such harmonics the intensity ratio between consecutive harmonics is 
$\frac{I_{2N}}{I_{2N+1}} \simeq (1\pm ee)^2 \frac{I_{2N}}{I_{2N+1}}\bigg|_{ee=0}$, where $\frac{I_{2N}}{I_{2N+1}}\bigg|_{ee=0}$ is the intensity ratio in a racemic mixture (see Fig. 3f and Methods).
The intensity of even harmonics is controlled at the level of total, integrated over emission angle, signal (Fig. \ref{fig_far}d). Such control is only possible because the handedness of the field is maintained globally in space.

One can also create locally chiral fields which carry several elements of chirality or change their helicity in space (see Supplementary Information). They may present alternative opportunities for studying matter with similar spatial patterns of chirality, or for exciting such patterns on demand.

The mechanism responsible for efficient control of chiral optical response with locally and globally chiral fields is general for isotropic chiral media in gases, liquids, solids, and plasmas.
Locally and globally chiral fields open new routes for chiral discrimination, enantio-sensitive imaging including selective monitoring of enantiomers with a specific handedness in non-enantiopure samples undergoing a chemical reaction, with ultrafast time resolution, laser-based separation of enantiomers, and efficient control of chiral matter.
They can also be used to imprint chirality on achiral matter, extending the recent proposal of Ref. \cite{Owens2018} from molecular rotations to various media and various degrees of freedom.  For example, (i) optical excitation of chiral electronic states in achiral systems such as the hydrogen atom\cite{ordonez2018propensity},
(ii) efficient enantiosensitive population of chiral vibronic states in molecules that are achiral in the ground state, e.g. formamide\cite{Rouxel2017SD},
or (iii) exciting chiral dynamics in delocalized systems that are not chiral, such as a free electron or, more generally, collective electronic excitations in metals and plasmas.
Looking broadly, these opportunities can be used to realize laser-driven ”achiral-chiral” phase transitions in matter.

As global control over the handedness of locally chiral fields creates enantio-sensitive time and space periodic structures, it 
may open interesting opportunities for exploring 
self-organization of quantum chiral matter and in the interaction of such fields with nano-structured chiral metamaterials.

\section*{METHODS}
\textbf{Chiral-field correlation functions.}
For a locally chiral field with polarization vector $\mathbf{F}(t)$ the lowest order chiral-field correlation function is a pseudoscalar
\begin{equation}
H^{(3)}(\tau_1,\tau_2)\equiv \frac{1}{2\pi}\int \mathrm{d}t\,\mathbf{F}(t)\cdot\left[\mathbf{F}(t+\tau_1)\times \mathbf{F}(t+\tau_2)\right]. 
\end{equation}
Its complex counterpart in the frequency domain,
\begin{equation}
h^{(3)}(-\omega_1-\omega_2,\omega_1,\omega_2)\equiv \int\mathrm{d}\tau_1 \int\mathrm{d}\tau_2\,H^{(3)}(\tau_1,\tau_2)e^{-i\omega_1\tau_1}e^{-i\omega_2\tau_2},
\end{equation}
yields Eq. \eqref{H3} of the main text.

$h^{(3)}$ describes perturbative enantio-sensitive three-photon light matter interaction.
For example, it appears as a light pseudoscalar in absorption circular dichroism in the electric-dipole approximation, analogously to how the helicity of circularly polarized light contributes to standard absorption circular dichroism beyond the electric-dipole approximation.

Consider a field with frequencies $\omega_1$, $\omega_2$, and $\omega_{+}\equiv \omega_2 + \omega_1$
\begin{equation}
\mathbf{F}\left(t\right)=\mathbf{F}_{1}e^{-i \omega_{1}t}+\mathbf{F}_{2}e^{-i \omega_{2}t}+\mathbf{F}_{+}e^{-i \omega_{+}t}+\mathrm{c.c.}
\label{field3_methods}
\end{equation}
The contribution from second order induced  polarization is enantio-sensitive\cite{Fischer2005} and can be written as (see also \cite{Giordmaine_1965}):
\begin{equation}
\mathbf{P}^{\left(2\right)}\left(\omega_{3}\right)=\frac{1}{3}\epsilon_{\alpha\beta\gamma}\chi_{\alpha\beta\gamma}^{\left(2\right)}\left(\omega_{3}=\omega_{1}+\omega_{2}\right)\left[\mathbf{F}\left(\omega_{1}\right)\times\mathbf{F}\left(\omega_{2}\right)\right].\label{eq:P2_general}
\end{equation}
Note that $\mathbf{P}^{(2)}(\omega_3)$ vanishes if $\omega_1=\omega_2$\cite{Fischer2005}.
Thus, for the field in Eq. (\ref{field3_methods}) we obtain:
\begin{equation}
\mathbf{P}^{\left(2\right)}\left(\omega_{1}\right)=\frac{1}{3}\epsilon_{\alpha\beta\gamma}\chi_{\alpha\beta\gamma}^{\left(2\right)}\left(\omega_{1}=\omega_{+}-\omega_{2}\right)\left(\mathbf{F}_{+}\times\mathbf{F}_{2}^{*}\right),\label{eq:P2_omega1}
\end{equation}
\begin{equation}
\mathbf{P}^{\left(2\right)}\left(\omega_{2}\right)=\frac{1}{3}\epsilon_{\alpha\beta\gamma}\chi_{\alpha\beta\gamma}^{\left(2\right)}\left(\omega_{2}=\omega_{+}-\omega_{1}\right)\left(\mathbf{F}_{+}\times\mathbf{F}_{1}^{*}\right),\label{eq:P2_omega_2}
\end{equation}
\begin{equation}
\mathbf{P}^{\left(2\right)}\left(\omega_{+}\right)=\frac{1}{3}\epsilon_{\alpha\beta\gamma}\chi_{\alpha\beta\gamma}^{\left(2\right)}\left(\omega_{+}=\omega_{2}+\omega_{1}\right)\left(\mathbf{F}_{2}\times\mathbf{F}_{1}\right).\label{eq:P2_omega_21}
\end{equation}
The second-order response at the difference frequency $\omega_2-\omega_1$ does not contribute to absorption because this frequency is absent in the driving field. 
Using the standard definition for  the total energy $\mathcal{E}$ exchanged between the field and the molecule,
\begin{eqnarray}
\mathcal{E} & = & \int_{-\infty}^{\infty}\mathrm{d}t \,\mathbf{F}\left(t \right)\cdot\dot{\mathbf{P}}\left(t \right)\nonumber\\
 & = & -2\pi i \int\mathrm{d}\omega\,\omega\mathbf{F}\left(-\omega\right)\cdot\mathbf{P}\left(\omega\right),
 \label{eq:absorption}
\end{eqnarray}
and replacing Eqs. \eqref{eq:P2_omega1}, \eqref{eq:P2_omega_2}, and \eqref{eq:P2_omega_21} in Eq. \eqref{eq:absorption} we obtain
\begin{multline}
\mathcal{E}^{\left(2\right)}
 = \frac{4\pi}{3}\mathrm{Im}\bigg\{\bigg[-\omega_{1}\epsilon_{\alpha\beta\gamma}\chi_{\alpha\beta\gamma}^{\left(2\right)*}\left(\omega_{1}=\omega_{+}-\omega_{2}\right)+\omega_{2}\epsilon_{\alpha\beta\gamma}\chi_{\alpha\beta\gamma}^{\left(2\right)*}\left(\omega_{2}=\omega_{+}-\omega_{1}\right)\\
 +\omega_{+}\epsilon_{\alpha\beta\gamma}\chi_{\alpha\beta\gamma}^{\left(2\right)}\left(\omega_{+}=\omega_{2}+\omega_{1}\right)\bigg]\left[\mathbf{F}_{+}^{*}\cdot\left(\mathbf{F}_{2}\times\mathbf{F}_{1}\right)\right]\bigg\}.
 \label{eq:enantioabsorption}
\end{multline}
Equation \eqref{eq:enantioabsorption} shows that enantio-sensitive light absorption is controlled by the third-order correlation function $h^{(3)}$ in Eq. \eqref{H3}. Indeed, it is proportional to the imaginary part of a product of two pseudoscalars: the first one is associated with chiral medium and involves second order susceptibilities, the second one is associated with the locally chiral field and is given by its correlation function $h^{(3)}=\left[\mathbf{F}_{+}^{*}\cdot\left(\mathbf{F}_{2}\times\mathbf{F}_{1}\right)\right]$. An analogous expression applies for a field with frequencies $\omega_1$, $\omega_2$, and $\omega_1 - \omega_2$.

\textbf{Chiral susceptibility tensor in Eqs. \eqref{interf} and \eqref{interf1}}.
Here we derive Eq. \eqref{interf} of the main text. The polarization corresponding to absorption of a single $2\omega$ photon is given by
\begin{equation}
    P_{i}^{\left(1\right)}\left(2\omega\right)=\chi_{ij}^{\left(1\right)}\left(2\omega \right)F_{j}\left(2\omega\right),
\end{equation}
where the orientation-averaged susceptibility is given by
\begin{eqnarray}
    \chi_{ij}^{\left(1\right)}&=&\left(\int\mathrm{d}\varrho\,l_{i\alpha}l_{j\beta}\right)\chi_{\alpha\beta}^{\left(1\right)}\nonumber\\
    &=&\frac{1}{3}\delta_{i,j}\delta_{\alpha,\beta}\chi_{\alpha\beta}^{\left(1\right)}
\end{eqnarray}
and yields
\begin{equation}
    \mathbf{P}^{(1)}(2\omega)=\frac{1}{3}\chi_{\alpha\alpha}^{\left(1\right)}\left(2\omega \right)\mathbf{F}\left(2\omega\right).
\end{equation}
Here $l_{i\alpha}$ is the direction cosine between axis $i$ in the lab frame and axis $\alpha$ in the molecular frame. We use latin indices for the lab frame and greek indices for the molecular frame.
The fourth-order term corresponding to absorption of three and emission of one $\omega$ photon reads as
\begin{equation}
    P_{i}^{\left(4\right)}\left(2\omega\right) = 4\chi_{ijklm}^{\left(4\right)}\left(2\omega =-\omega+\omega+\omega+\omega\right)F_{j}^{*}\left(\omega\right)F_{k}\left(\omega\right)F_{l}\left(\omega\right)F_{m}\left(\omega\right),
\end{equation}
where the degeneracy factor 4 comes from the four possible photon orderings. The orientation-averaged fourth-order susceptibility in the lab frame is given by
\begin{equation}
    \chi_{ijklm}^{(4)} = \left(\int \mathrm{d} \varrho\,l_{i\alpha}l_{j\beta}l_{k\gamma}l_{l\delta}l_{m\epsilon} \right) \chi_{\alpha\beta\gamma\delta\epsilon}.
\end{equation}
Using standard expressions for the orientation averaging (see e.g. Refs. \cite{Ordonez2018_PRA} and \cite{Andrews_1977}) we obtain (see also\cite{Fischer2005}):
\begin{multline}
    \mathbf{P}^{\left(4\right)}\left(2\omega \right) =\frac{2}{15}\chi_{\alpha\beta\gamma\delta\epsilon}^{\left(4\right)}\left(2\omega =-\omega+\omega+\omega+\omega\right)\\
    \times\left(\epsilon_{\alpha\beta\gamma}\delta_{\delta\epsilon}+\epsilon_{\alpha\beta\delta}\delta_{\gamma\epsilon}+\epsilon_{\alpha\beta\epsilon}\delta_{\gamma\delta}\right)\left[\mathbf{F}^{*}\left(\omega\right)\times\mathbf{F}\left(\omega\right)\right]\left[\mathbf{F}\left(\omega\right)\cdot\mathbf{F}\left(\omega\right)\right].
\end{multline}
In Eq. \eqref{interf} we have used  shorthand notations for the first-order and the fourth-order susceptibilities, i.e.
\begin{equation}
    \chi^{(1)} \equiv \frac{1}{3}\chi_{\alpha\alpha}^{\left(1\right)}\left(2\omega \right),
\end{equation}
\begin{equation}
    \chi^{(4)}_{\pm} \equiv \frac{2}{15}\chi_{\alpha\beta\gamma\delta\epsilon}^{\left(4\right)}\left(2\omega =-\omega+\omega+\omega+\omega\right)\left(\epsilon_{\alpha\beta\gamma}\delta_{\delta\epsilon}+\epsilon_{\alpha\beta\delta}\delta_{\gamma\epsilon}+\epsilon_{\alpha\beta\epsilon}\delta_{\gamma\delta}\right).
\end{equation}
Note that for a field containing only frequencies $\omega$ and $2\omega$ the second-order response at frequency $2\omega$ vanishes [see Eq. \eqref{eq:P2_general}] and therefore there is no interference between $\mathbf{P}^{(2)}(2\omega)$ and $\mathbf{P}^{(3)}(2\omega)$. Although $P^{(3)}(2\omega)$ is non-zero, it behaves either as $\vert F(\omega) \vert^{2}F(2\omega)$ or as $\vert F(2\omega)\vert^{2}F(2\omega)$. The last term should be omitted, since we keep only terms linear in $F(2\omega)$. As for the first term, which includes additional absorption and emission of $\omega$ photons, it merely describes the standard nonlinear modification of the linear response to the weak $2\omega$ field due to the polarization of the system by the strong $\omega$ field, leading e.g. to the Stark shifts of the states involved. Thus, while these terms do modify the effective linear susceptibility $\chi^{(1)}(2\omega)$, which should include the dressing of the system by the strong $\omega$ field, they do not modify the overall result.



\textbf{Fifth-order chiral-field correlation function}. 
The different chiral-field correlation functions $h^{(n)}$, with $n$ odd, control the sign of enantio-sensitive and dichroic response in multiphoton interactions. For the locally chiral field employed in our work [see Eq. \eqref{eq_field}] $h^{(5)}$ is the lowest-order non-vanishing chiral-field correlation function. Here we show that $h^{(5)}$ has a unique form for this field, as may be anticipated from the derivation in the previous section. 

We write the field in Eq. \eqref{eq_field} in the exponential form
\begin{equation}
\boldsymbol{F}\left(t\right) =  \left(F_{x}\hat{\boldsymbol{x}}+ i F_{y}\hat{\boldsymbol{y}}\right)e^{ -i\left(\omega t+\delta_{\omega}\right)}+F_{z} \hat{\boldsymbol{z}}e^{ -2i\left(\omega t+\delta_{2\omega}\right)}+\mathrm{c.c.},\label{eq:field_h5}
\end{equation}
and the fifth-order chiral-field correlation function as
\begin{equation}
h^{(5)}(\omega_0,\omega_1,\omega_2, \omega_3,\omega_4)\equiv \left\{ \boldsymbol{F}(\omega_0)\cdot[\boldsymbol{F}(\omega_1)\times \boldsymbol{F}(\omega_2)]\right\}\left[\boldsymbol{F}(\omega_3)\cdot\boldsymbol{F}(\omega_4)\right],\label{eq:h5_methods}
\end{equation}
where 
\begin{equation}
\omega_0+\omega_1+\omega_2+\omega_3+\omega_4=0.\label{eq:energy_conservation}
\end{equation}
First, note that $h^{\left(5\right)}$ is symmetric with respect to
exchange of $\omega_{3}$ and $\omega_{4}$, and symmetric/anti-symmetric
with respect to even/odd permutations of $\omega_{0}$, $\omega_{1}$,
and $\omega_{2}$. Non-trivially different forms of $h^{\left(5\right)}$
result only from considering exchanges between $\left\{ \omega_{0},\omega_{1},\omega_{2}\right\} $
and $\left\{ \omega_{3},\omega_{4}\right\} $. In the following we show
that the field (Eq. \eqref{eq_field} of the main text) yields a unique non-zero form of $h^{\left(5\right)}$. 

For $h^{\left(5\right)}$ to be non-vanishing, the triple product
in $h^{\left(5\right)}$ must contain the three different vectors
available in our field, namely $F_{x}\hat{x}+ i F_{y}\hat{y}$, $F_{x}\hat{x}-i F_{y}\hat{y}$
(from the c.c. part), and $\hat{z}$. The remaining scalar product
must have frequencies such that Eq. \eqref{eq:energy_conservation} is satisfied.
This means that we have the following four options for $h^{\left(5\right)}$:
\begin{eqnarray}
h_{a}^{\left(5\right)} & = & h^{\left(5\right)}\left(-2\omega,-\omega,\omega,\omega,\omega\right)\label{eq:ha} \\
h_{b}^{\left(5\right)} & = & h^{\left(5\right)}\left(-2\omega,\omega,-\omega,\omega,\omega\right)\\
h_{c}^{\left(5\right)} & = & h^{\left(5\right)}\left(2\omega,\omega,-\omega,-\omega,-\omega\right)\\
h_{d}^{\left(5\right)} & = & h^{\left(5\right)}\left(2\omega,-\omega,\omega,-\omega,-\omega\right)\label{eq:hd}
\end{eqnarray}
Since $\vec{F}(\omega_2)\times\vec{F}(\omega_1)=-\vec{F}(\omega_1)\times\vec{F}(\omega_2)$ and $\vec{F}(t)$ is real, then
\begin{equation}
h_{b}^{\left(5\right)}=-h_{a}^{\left(5\right)}
\end{equation}
\begin{equation}
h_{c}^{\left(5\right)}=h_{a}^{\left(5\right)*}
\end{equation}
\begin{equation}
h_{d}^{\left(5\right)}=-h_{c}^{\left(5\right)}\label{eq:hd_hc}
\end{equation}
so that all options are actually equivalent to each other up to either
a sign, a complex conjugation, or both. The ambiguity in the sign
reflects the fact that in general right-handed and left-handed cannot
be defined in an absolute way, i.e. without reference to another chiral
object, so it is equally valid to define a chiral measure $h^{(5)}=h_{a}^{(5)}$
or a chiral measure $h^{(5)}=h_{b}^{5)}=-h_{a}^{(5)}$ to
characterize the ``absolute'' handedness of the field.
Choosing $h^{(5)}_a$ or $h^{(5)}_b$, i.e. ordering vectors in the vector product, corresponds to choosing a left- or a right-handed reference frame.
On the other hand, if we consider the interaction with matter, $h^{\left(5\right)}$ and $h^{\left(5\right)*}$
appear on the same footing [see Eq. \eqref{interf}], therefore
it is only natural that we have found both options in this derivation.
Even if we consider the field by itself, since $h^{\left(5\right)}$
is the Fourier transform of a real quantity, $h^{\left(5\right)}$
and $h^{\left(5\right)*}$ contain the same information, so it is
again to be expected that we find both options in our derivation.

Replacing Eqs. \eqref{eq:field_h5} and \eqref{eq:h5_methods} in Eq. \eqref{eq:ha} we obtain
\begin{equation}
h_{a}^{\left(5\right)}=2 i F_{x}F_{y}F_{z}\left(F_{x}^{2}-F_{y}^{2}\right) e^{ 2i\left(\delta_{2\omega}-\delta_{\omega}\right)},\label{eq:h5_a}
\end{equation}
so that $h_{a}^{\left(5\right)}$ is sensitive to the phase difference
between the two colors and also to the sign of the ellipticity of
the $\omega$ field $\varepsilon\equiv F_{y}/F_{x}$. 
In the supplementary information we show that, although the chiral-field correlation function $h^{(n)}$ does not have a unique non-zero form for higher odd orders $n=7,9,...$, only a single form is non-negligible provided that the fields along $y$ and $z$ are weak in comparison to the field along $x$, such as the field employed in the main text.

\textbf{Experimental realization of locally chiral fields.}
The locally chiral field shown in Fig. \ref{fig_scheme}a can be realized using the non-collinear setup presented in Fig. \ref{fig_near}a.
It consists of two beams that propagate along $\mathbf{k}_1$ and $\mathbf{k}_2$, in the $xy$ plane, creating angles $\pm\alpha$ with the $y$ axis.
Each beam is composed of a linearly polarized field, whose polarization is contained in the $xy$ plane, and a $z$-polarized second harmonic.
We shall assume that the two beams ($n=1$, $2$) are Gaussian beams\cite{book_Boyd}, and thus their electric fields can be written at the focus as
\begin{align}
\mathbf{F}_n^{ \omega}(\mathbf{r},t) &=     F_0 \hspace{0.2em} e^{-\rho_n^2/\tilde{\omega}^2} \cos{(\mathbf{k}_n\cdot\mathbf{r} -  \omega t - \phi_n^{ \omega})} \hspace{0.2em}  \hat{\mathbf{e}}_n \label{eq_Fw_focus} \\
\mathbf{F}_n^{2\omega}(\mathbf{r},t) &= r_0 F_0 \hspace{0.2em} e^{-\rho_n^2/\tilde{\omega}^2} \cos{(2\mathbf{k}_n\cdot\mathbf{r} - 2\omega t - 2\phi_n^{2\omega})} \hspace{0.2em}  \hat{\mathbf{z}}   \label{eq_F2w_focus}
\end{align}
where 
$F_0$ is the electric field amplitude,
$r_0^2$ is the intensity ratio between the two colours,
$\rho_n$ is the radial distance to beams' axis ($\rho=\rho_1\simeq\rho_2$ in the focus), 
$\tilde{\omega}$ is the waist radius,
the propagation vectors of the fundamental field are defined as
\begin{align}
\mathbf{k}_1 &=  k\sin{(\alpha)}\hat{\mathbf{x}} + k\cos{(\alpha)}\hat{\mathbf{y}} \\
\mathbf{k}_2 &= -k\sin{(\alpha)}\hat{\mathbf{x}} + k\cos{(\alpha)}\hat{\mathbf{y}}
\end{align}
where $k=\frac{2\pi}{\lambda}$, $\lambda$ being the fundamental wavelength,
and the polarization vectors are given by
\begin{align}
\hat{\mathbf{e}}_1 = \cos{(\alpha)}\hat{\mathbf{x}} - \sin{(\alpha)}\hat{\mathbf{y}} \\
\hat{\mathbf{e}}_2 = \cos{(\alpha)}\hat{\mathbf{x}} + \sin{(\alpha)}\hat{\mathbf{y}}
\end{align}
The total electric field can be written as
\begin{align}
\mathbf{F}(\mathbf{r},t) = 2 F_0 \hspace{0.2em} e^{-\rho^2/\tilde{\omega}^2} \Big[ &f_x(x) \cos{\big( k\cos{(\alpha)}y- \omega t-\delta_{+}^{\omega}\big)}    \hat{\mathbf{x}} + f_y(x) \sin{\big(k\cos{(\alpha)}y-\omega t-\delta_{+}^{\omega}\big)} \hat{\mathbf{y}} \nonumber \\
                                                                             + &f_z(x) \cos{\big(2k\cos{(\alpha)}y-2\omega t-2\delta_{+}^{2\omega}\big)} \hat{\mathbf{z}} \Big]     \label{eq_F_focus}
\end{align}
with
\begin{align}
f_x(x) &=     \cos{(\alpha)} \cos{\Big(  k\sin{(\alpha)}x+\delta_{-}^{ \omega} \Big)} \label{eq_Fx_focus} \\ 
f_y(x) &=     \sin{(\alpha)} \sin{\Big(  k\sin{(\alpha)}x+\delta_{-}^{ \omega} \Big)} \label{eq_Fy_focus} \\
f_z(x) &= r_0                \cos{\Big( 2k\sin{(\alpha)}x+2\delta_{-}^{2\omega} \Big)} \label{eq_Fz_focus}
\end{align}
where $\delta_{\pm}^{m\omega} = (\phi_2^{m\omega} \pm \phi_{1}^{m\omega})/2$.
Equations \eqref{eq_F_focus}-\eqref{eq_Fz_focus} show that the total electric field is locally chiral.
It is elliptically polarized in the $xy$ plane at frequency $\omega$ (with major polarization component along $x$) and linearly polarized along $z$ at frequency $2\omega$.

Note that the relative phases between the three field components do not change along $z$, as the two beams propagate in the $xy$ plane.
They do not change along $y$ either, since $\mathbf{k_1}\cdot\mathbf{y}=\mathbf{k_2}\cdot\mathbf{y}$.
Indeed, one can easily see in Eq. \eqref{eq_F_focus} that a spatial translation $y\rightarrow y+\Delta y$ is equivalent to a temporal displacement $t\rightarrow t-\Delta t$, with $k\cos({\alpha})\Delta y = \omega\Delta t$.
However, the relative phases between the field components do change along $x$, because $\mathbf{k_1}\cdot\mathbf{x}\neq\mathbf{k_2}\cdot\mathbf{x}$ (we have $\mathbf{k_1}\cdot\mathbf{x}=-\mathbf{k_2}\cdot\mathbf{x}$ instead), and their modulation is given by $f_x(x)$, $f_y(x)$ and $f_z(x)$.

In order to generate a macroscopic chiral response, we need to ensure that the \emph{handedness} of the locally chiral field is locked throughout space. 
As $f_x$ and $f_y$ change along $x$, so does the ellipticity in the $xy$ plane, which can be defined as
\begin{equation}\label{eps_f}
\varepsilon(x) = \frac{f_y(x)}{f_x(x)} = \tan{(\alpha)} \tan{(k\sin{(\alpha)}x+\delta_{-}^{\omega})} 
\end{equation}
One can easily see that the field's \emph{handedness} depends on the relative sign between $\varepsilon(x)$ and $f_z(x)$.
Thus, we just need to make sure that both quantities change sign at the same positions.
The spatial points where forward ellipticity flips sign satisfy
\begin{equation}\label{eq_condition_eps}
k\sin{(\alpha)}x +\delta_{-}^{\omega} = n\frac{\pi}{2}, \quad \text{with } n\in\mathbb{Z}.
\end{equation}
Whereas for $f_z$ we have
\begin{equation}\label{eq_condition_fx}
2k\sin{(\alpha)}x + 2\delta_{-}^{2\omega} = \frac{\pi}{2}+n\pi, \quad \text{with } n\in\mathbb{Z}.
\end{equation}
Combining Eqs. \eqref{eq_condition_eps} and \eqref{eq_condition_fx}, we obtain the following general condition
\begin{equation}\label{eq_condition_phases}
2\delta_{-}^{\omega} - 2\delta_{-}^{2\omega} = \frac{\pi}{2} + n\pi, \quad \text{with } n\in\mathbb{Z}
\end{equation}
Let us consider now the situation where the two fundamental fields are out of phase (as in Fig. \ref{fig_near}a), i.e. $\phi_1^{\omega}=\delta_{+}^{\omega}-\pi/2$ and $\phi_2^{\omega}=\delta_{+}^{\omega}+\pi/2$, and therefore $\delta_{-}^{\omega}=\pi/2$.
Then, the condition given by Eq. \eqref{eq_condition_phases} is verified if $2\delta_{-}^{2\omega}=\pi/2$, i.e. if the second harmonic fields are also out of phase, and then we have $2\phi_1^{2\omega}=2\delta_{+}^{2\omega}-\pi/2$ and $2\phi_2^{2\omega}=2\delta_{+}^{2\omega}+\pi/2$.
This means that the relative phase between the two colours has to be the same in both beams, $2\phi_n^{2\omega}-\phi_n^{\omega}=2\delta_{+}^{2\omega}-\delta_{+}^{\omega}$.
This analysis shows that the locally chiral field shown in Fig 1a of the main text 
maintains its handedness globally in space.
It can also be seen from the chiral-field correlation functions.

\textbf{Global handedness in chiral-field correlation functions}
Here we analyze how the handedness of the chiral field Eq. \eqref{eq_field} as defined via the chirality measure $h^{(5)}$ in Eq. \eqref{h5_even} changes in space. From Eqs. \eqref{eq:field_h5}, \eqref{eq:h5_a}, and \eqref{eq_F_focus}-\eqref{eq_Fz_focus}, we can see that $h^{(5)}$ depends on $x$ through $F_x$, $F_y$, and $F_z$.
$F_x$ and $F_y$ oscillate with a frequency $k\sin\alpha$, and $F_z$ oscillates with a frequency $2k\sin\alpha$, as a function of $x$. Of course, $h^{(5)}$ also oscillates as a function of $x$, however, by decomposing $F_x$, $F_y$, and $F_z$ into exponentials with positive and negative frequencies, one can see that there is also a null-frequency component in $h^{(5)}$ that does not oscillate as a function of $x$. Ultimately, it is this null-frequency component which defines the global handedness of the field. It is given by
\begin{eqnarray}
\left[h_{a}^{(5)}\right]_{0} &=& 2i\left[F_{x}F_{y}F_{z}\left(F_{x}^{2}-F_{y}^{2}\right)\right]_{0} e^{2i\left(\delta_{+}^{2\omega} - \delta_{+}^{\omega}\right)} \nonumber\\
&=&  \frac{1}{8}F_{0}^{5}r_0\sin\left(4\alpha\right)\sin\left(2\delta_{-}^{2\omega} - 2\delta_{-}^{\omega}\right) e^{i\left(2\delta_{+}^{2\omega} - 2\delta_{+}^{\omega}+\frac{\pi}{2}\right)}. \label{eq:h5_0frequency_in_x}
\end{eqnarray}
This expression shows that the global handedness of the field vanishes when the relative phases satisfy $2\delta_{-}^{2\omega}-2\delta_{-}^{\omega}=n\pi$ for integer $n$. On the other hand, the absolute value of the global handedness reaches a maximum when $2\delta_{-}^{2\omega}-2\delta_{-}^{\omega}=(n+1)\pi/2$ for integer $n$, in accordance with Eq. \eqref{eq_condition_phases}. This condition is satisfied by the field shown in  Fig. 2 of the main text. 

\textbf{High harmonic response in propylene oxide.}
We have adapted the method described in Ref. \cite{bookChapter_SmirnovaIvanov_AttosecondAndXUVPhysics} to describe high harmonic response in the chiral molecule propylene oxide, as in\cite{Ayuso2018JPB}.
The macroscopic dipole driven in a medium of randomly oriented molecules results from the coherent addition of all possible molecular orientations, i.e.
\begin{equation}\label{eq_dipole}
\textbf{D}(N\omega) = \int d\Omega \int d\alpha \hspace{0.2em} \textbf{D}_{\Omega\alpha}(N\omega)
\end{equation}
where where $\omega$ is the fundamental frequency,
$N$ is the harmonic number,
and $\textbf{D}_{\Omega\alpha}$ is the harmonic dipole associated with a given molecular orientation.
The integration in the solid angle $\Omega$ was performed using the Lebedev quadrature\cite{Lebedev1999} of order 17.
For each value of $\Omega$, the integration in $\alpha$ was evaluated using the trapezoid method.

The contribution from each molecular orientation results from the coherent addition of all channel contributions\cite{bookChapter_SmirnovaIvanov_AttosecondAndXUVPhysics}:
\begin{equation}
\textbf{D}_{\Omega\alpha}^{mn}(N\omega) = \sum_{mn}\textbf{D}_{mn}(N\omega)
\end{equation}
where $\textbf{D}_{mn}$ is the contribution from a given ionization ($m$) - recombination ($n$) channel, in the frequency domain.
We have considered the electronic ground state of the ionic core (X) and the first three excited states (A, B and C), i.e. 16 channels,
and found that only those involving ionization from the X and A states and recombination with the X, A and B states play a key role under the experimental conditions of Ref. \cite{Cireasa2015NatPhys}.
The contribution from a single ionization-recombination burst can be factorized as a product of three terms:
\begin{equation}
\textbf{D}_{\Omega\alpha}^{mn}(N\omega) = a_{ion,\Omega\alpha}^{mn}(N\omega) \cdot a_{prop,\Omega\alpha}^{mn}(N\omega) \cdot \textbf{a}_{rec,\Omega\alpha}^{mn}(N\omega)
\end{equation}
which are associated with strong-field ionization, propagation and radiative recombination, respectively\cite{bookChapter_SmirnovaIvanov_AttosecondAndXUVPhysics}.

Recombination amplitudes are given by
\begin{equation}
\textbf{a}_{rec,\Omega\alpha}^{nm} = \bigg(\frac{2\pi}{i\partial^2 S_m(t_r,t_i,\textbf{p})/\partial t_r^2}\bigg)^{1/2} \hspace{0.2em} e^{-iS_m(t_r,t_r',\textbf{p})+iN\omega t_r} \hspace{0.2em} \textbf{d}_{rec,n}^{\Omega\alpha}\big(\textbf{k}(t_r')\big)
\end{equation}
where $t_i=t_i'+it_i''$ and $t_r=t_r'+it_r''$ are the complex ionization and recombination times resulting from applying the saddle-point method\cite{bookChapter_SmirnovaIvanov_AttosecondAndXUVPhysics},
$\textbf{p}$ represents the canonical momentum, which is related to the kinetic momentum by $\textbf{k}(t)=\textbf{p}(t)+\textbf{A}(t)$, $\textbf{A}(t)$ being the vector potential ($\textbf{F}(t)=-\partial\textbf{A}(t)/\partial{t}$),
$\textbf{d}_{rec,n}$ is the corresponding photorecombination matrix element,
and $S_m$ is given by
\begin{equation}\label{eq_VolkovPhase}
S_m(t,t',\textbf{p}) = \frac{1}{2} \int_{t'}^{t} d\tau \big[\textbf{p}+\textbf{A}(\tau)\big]^2 + \textbf{IP}_{m} (t-t')
\end{equation}
Photorecombination matrix elements have been evaluated the static-exchange density functional theory (DFT) method\cite{Toffoli2002CP,Bachau2001RPP,Turchini2004PRA,StenerJCP2004,Stranges2005JCP}, as in\cite{Ayuso2018JPB}.

Propagation amplitudes are given by
\begin{equation}
a_{prop,\Omega\alpha}^{nm} = \bigg(\frac{2\pi}{i(t_r-t_i)}\bigg)^{3/2}  e^{-iS_m(t_r',t_i',\textbf{p})} a_{mn}^{\Omega\alpha}(t_r',t_i')
\end{equation}
where $a_{mn}^{\Omega\alpha}$ is the transition amplitude describing the laser-electron dynamics between ionization and recombination, which is obtained by solving the time-dependent Schr\"odinger equation numerically in the basis set of ionic states\cite{bookChapter_SmirnovaIvanov_AttosecondAndXUVPhysics}.

A reasonable estimation of the ionization amplitudes can be obtained using the following expression:
\begin{equation}\label{eq_ionization}
a_{ion,\Omega\alpha}^{nm} = 2\pi \bigg(\frac{1}{i\partial^2 S_m(t_r,t_i,\textbf{p})/\partial t_i^2}\bigg)^{1/2} e^{-iS_m(t_i',t_i,\textbf{p})} \mathcal{F}\{\Psi_m\}\big(\Re\{\textbf{k}(t_i')\}\big)
\end{equation}
where $\mathcal{F}\{\Psi_m\}$ is the Fourier transform of the Dyson orbital associated with the initial state wave function $\Psi_m$.
The evaluation of sub-cycle ionization amplitudes in organic molecules is very challenging because non-adiabatic and multi-electron effects influence the dynamics of laser-driven electron tunneling, and the estimations provided by Eq. \eqref{eq_ionization} are not sufficiently accurate for the purpose of this work.
Nonetheless, these quantities can be reconstructed from multi-dimensional HHG spectroscopy measurements, when available\cite{Serbinenko2013JPB,Pedatzur2015,Bruner2016FD}.
Here we have reconstructed the amplitudes and phases of the sub-cycle ionization amplitudes from the experimental results of Ref. \cite{Cireasa2015NatPhys}, using the estimations provided by Eq. \eqref{eq_ionization} as a starting point for the procedure.

\textbf{Evaluation of macroscopic chiral response}
The harmonic intensity in the far field has been evaluated using the Fraunhofer diffraction equation, i.e.
\begin{equation}\label{Fraunhofer}
\mathbf{I}(N\omega,\beta) \propto (N\omega)^4 \hspace{0.2em} \Bigg| \int_{-\infty}^{\infty} \mathbf{D}(N\omega,x) e^{-iK x}  dx \Bigg|^2
\end{equation}
where $\beta$ is the far field angle (divergence),
and $K$ is given by $K=\frac{N\omega}{c}\beta$, $c$ being the speed of light,
and $\mathbf{D}(N\omega,x)$ is the harmonic dipole driven by the strong field in the focus (Eq. \eqref{eq_dipole}), which has been computed along the transversal coordinate $x$ using the procedure described in the previous section.

\textbf{Accurate determination of the enantiomeric excess.}
Let us consider a macroscopic mixture of right handed and left handed molecules, with concentrations $C_R$ and $C_S$, respectively.
The intensity of the odd harmonics, which is not enantiosensitive in the dipole approximation, depends on the total concentration of molecules, being proportional to
\begin{equation}
I_{2N+1} \propto (C_R+C_S)^2 |D_{x,2N+1}|^2
\end{equation}
where $D_{x,2N+1}$ is dipole component along $x$, averaged over all molecular orientations.
This component is essentially driven by the $x$ component of the fundamental field.
In other words, the dominant pathway giving rise to emission of even harmonics consists of the absorption of $2N+1$ photons with frequency $\omega$ and $x$ polarization. 
Thus, $I_{2N+1}$ is essentially unaffected by the presence of the second harmonic field, provided its intensity is weak.

Even harmonics are polarized along $z$, and they result from the interference between the chiral and achiral pathways depicted in Fig. \ref{fig_SF_diagrams} of the SI.
The intensity of even harmonics is given by
\begin{equation}
I_{2N} \propto | (C_R+C_S) D_{z,2N}^{0} + (C_R-C_S) D_{z,2N}^{(R)}|^2
\end{equation}
where $D_{z,2N}^{0}$ is the non-enantiosensitive dipole component associated with the achiral pathways depicted in Fig. \ref{fig_SF_diagrams} of the SI and $D_{z,2N}^{(R)}$ is the enantiosensitive component associated with the chiral pathway, for the $R$ enantiomer.
The strength of chiral response depends on $(C_R-C_S)$ because the enantiosensitive dipole component is out of phase in opposite enantiomers, i.e. $D_{z,2N}^{(R)}=-D_{z,2N}^{(S)}$.
The effect of the second harmonic field on $D_{z,2N}^{(R)}$ is negligible, as this dipole component is driven by the fundamental.
However, the achiral pathways giving rise to $D_{z,2N}^{0}$ involve the absorption or emission of one $z$-polarized photon of frequency $2\omega$, and thus this dipole component is controlled by the amplitude and phase of the second harmonic field.
As we show in the main text, we can tune these parameters so that $D_{z,2N}^{0}\simeq\pm D_{z,2N}^{(R)}$.
If $D_{z,2N}^{0}\simeq D_{z,2N}^{(R)}$, the ratio between consecutive harmonics can be written as
\begin{equation}\label{eq_determination_ee_1}
\frac{I_{2N}}{I_{2N+1}} \propto (1+ee)^2 \frac{I_{2N}}{I_{2N+1}}\Big|_{ee=0}
\end{equation}
where $ee$ is the enantiomeric excess, $ee = \frac{C_R-C_S}{C_R+C_S}$, and $\frac{I_{2N}}{I_{2N+1}}\Big|_{ee=0}$ is the intensity ratio between consecutive harmonics in a racemic mixture.
Alternatively, one can adjust the amplitude and phase of the second harmonic field so that $D_{z,2N}^{0} \simeq D_{z,2N}^{(S)}$, and then
\begin{equation}\label{eq_determination_ee_2}
\frac{I_{2N}}{I_{2N+1}} \propto (1-ee)^2 \frac{I_{2N}}{I_{2N+1}}\Big|_{ee=0}
\end{equation}
Eqs. \eqref{eq_determination_ee_1} and \eqref{eq_determination_ee_2} provide an easy way to quantify the enantiomeric excess in macroscopic mixtures from a single measurement of the harmonic spectrum, with high accuracy and with sub-femtosecond time resolution.
Note that, if $ee>0$, we can determine its value more accurately if we set $D_{z,2N}^{0} \simeq D_{z,2N}^{(R)}$ and use Eq. \eqref{eq_determination_ee_1},
whereas setting $D_{z,2N}^{0} \simeq D_{z,2N}^{(S)}$ and using Eq. \eqref{eq_determination_ee_2} will provide a more accurate determination if $ee<0$ (see Fig. \ref{fig_far}f).

\newpage
\section*{SUPPLEMENTARY INFORMATION}

\section{Benchmark of quantitative model for high harmonic response in propylene oxide}

\begin{figure}
\centering
\includegraphics[width=\linewidth, keepaspectratio=true]{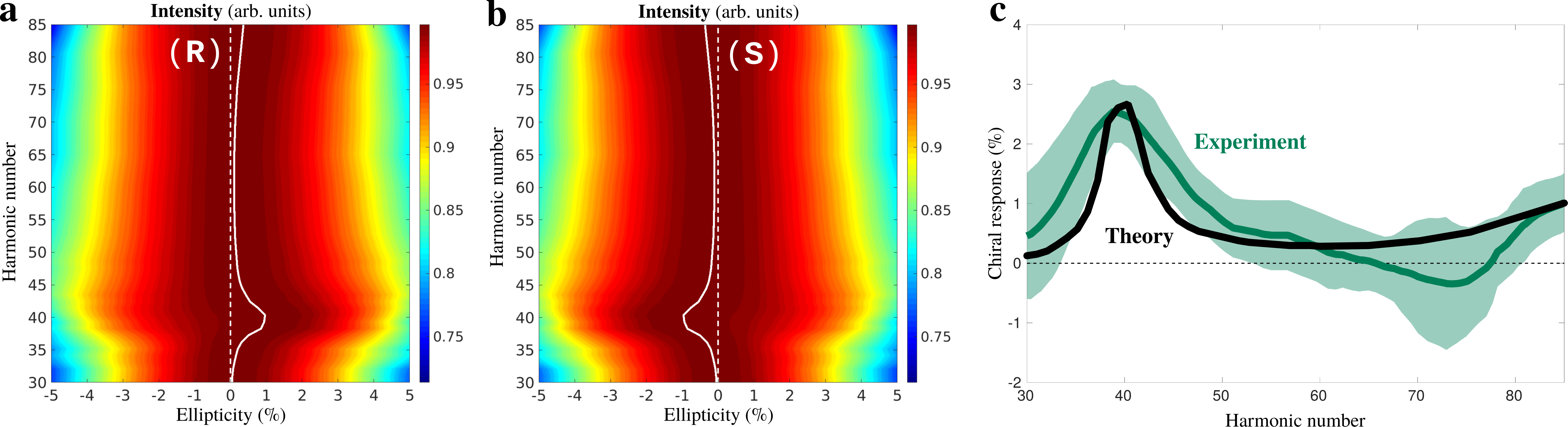}
\caption{a, b): High-order harmonic intensity emitted by randomly oriented $R$ and $S$ propylene oxide molecules in elliptically polarized laser fields with intensity $I_0=5\cdot10^{13}$ W cm$^{-2}$ and wavelength $\lambda=1770$ nm (see Methods for detail of the calculations).
For each harmonic number, the intensity is normalized to its maximum value.
The values of ellipticity that maximize the harmonic intensity are represented with a white line.
c) Time-resolved chiral response: theoretical results of this work (black line) and experimental values of Ref. \cite{Cireasa2015NatPhys} (green line).
The shaded area represents the uncertainty of the experimental measurements.
}
\label{fig_benchmark}
\end{figure}

The quantitative model for evaluating high harmonic response in propylene oxide (see Methods) used in this work has been benchmarked against the experimental results of Ref.\cite{Cireasa2015NatPhys}, which recorded the harmonic emission from randomly oriented molecules in elliptically polarized laser fields.
Fig. \ref{fig_benchmark}a,b contains the calculated high harmonic intensity for right-handed and left handed propylene oxide molecules as functions of harmonic number and ellipticity.
The agreement with the experimental results of Ref.\cite{Cireasa2015NatPhys} is excellent (see Fig. 1 in Ref.\cite{Cireasa2015NatPhys}).
Our quantitative model reproduces very well the chiral response around H40 and near the cutoff, which could not be explained within the simplified picture used in \cite{Cireasa2015NatPhys}.
Note that the origin of chiral response in elliptical HHG relies on the interplay between electric and magnetic dipole interactions, and thus it is not very strong (around $2-3\%$).
The use of locally chiral fields can enhance chiral response by two orders of magnitude, as we show in the main text.

Chiral response in HHG driven by weakly elliptically polarized fields is proportional to the ellipticity that maximizes harmonic signal\cite{Cireasa2015NatPhys}, and it is given by
\begin{equation}\label{eq_S}
S(N) \simeq 2 \frac{\varepsilon_{0}(N)}{\sigma^2}
\end{equation}
where $\varepsilon_{0}$ is the value of ellipticity that maximizes the harmonic signal, for a given harmonic number $N$, and $\sigma$ describes the Gaussian decay of the harmonic signal with ellipticity.
The values of $S(N)$ evaluated using the numerical results presented in Fig. \ref{fig_benchmark} (a,b) are shown in Fig. \ref{fig_benchmark}c, together with the experimental results from Ref. \cite{Cireasa2015NatPhys}.
The agreement between theory and experiment is excellent in the whole range of harmonic numbers, i.e. for all recombination times.

\section{Higher order chiral-field correlation functions}
Higher order chiral-field correlation functions $h^{(n)}$ control the sign of the enantio-sensitive and dichroic response in multiphoton interactions. Here, we consider higher-order ($n>5$) correlation functions $h^{(n)}$ for the locally chiral field Eq. (\ref{eq_field}) employed in our work. We show that for such field $h^{(n)}$ has a unique form in every order, which is helpful for achieving the ultimate control. 

The $n$-th order chiral correlation function in the time domain is defined as
\begin{multline}
H^{\left(n\right)}\left(0,\tau_{1},\dotsc,\tau_{n-1}\right)\equiv\frac{1}{\sqrt{2\pi}}\int\mathrm{d}t\,\left\{ \boldsymbol{F}\left(t\right)\cdot\left[\boldsymbol{F}\left(t+\tau_{1}\right)\times\boldsymbol{F}\left(t+\tau_{2}\right)\right]\right\}\\
\times\left[\boldsymbol{F}\left(t+\tau_{3}\right)\cdot\boldsymbol{F}\left(t+\tau_{4}\right)\right]\dotsc\left[\boldsymbol{F}\left(t+\tau_{n-2}\right)\cdot\boldsymbol{F}\left(t+\tau_{n-1}\right)\right]
\end{multline}
for $n\geq3$ odd. The Fourier transform with respect to all variables $\tau_i$ yields the $n$-th order chiral correlation function in the frequency domain:
\begin{eqnarray}
h^{\left(n\right)}\left(\omega_{0},\omega_{1},\dotsc,\omega_{n-1}\right) & = & \frac{1}{\left(\sqrt{2\pi}\right)^{n}}\int\mathrm{d}\tau_{1}\dotsc\int\mathrm{d}\tau_{n-1}e^{i\omega_{1}\tau_{1}}\cdots e^{i\omega_{n-1}\tau_{n-1}}H^{\left(n\right)}\left(0,\tau_{1},\dotsc,\tau_{n-1}\right)\nonumber\\
 & = & \left\{ \boldsymbol{F}\left(\omega_{0}\right)\cdot\left[\boldsymbol{F}\left(\omega_{1}\right)\times\boldsymbol{F}\left(\omega_{2}\right)\right]\right\}\nonumber\\
 &   & \times\left[\boldsymbol{F}\left(\omega_{3}\right)\cdot\boldsymbol{F}\left(\omega_{4}\right)\right]\dotsc\left[\boldsymbol{F}\left(\omega_{n-2}\right)\cdot\boldsymbol{F}\left(\omega_{n-1}\right)\right]
\end{eqnarray}
where $\omega_0$ is defined by the equation
\begin{equation}
\sum_{i=0}^{n-1}\omega_i=0
\end{equation}
We now consider all possible permutations of the frequencies in $h^{(n)}$, which is equivalent to considering all possible permutations of times in $H^{(n)}$. We will show that the handedness of the field employed in our work [see Eq. \eqref{eq:field_h5} in Methods] is invariant with respect to such permutations.

The seventh-order chiral-field correlation function reads as:
\begin{multline}
h^{\left(7\right)}\left(\{\omega_{0},\omega_{1},\omega_{2}\},[\omega_{3},\omega_{4}],[\omega_{5}, \omega_{6}]\right) = \left\{ \boldsymbol{F}\left(\omega_{0}\right)\cdot\left[\boldsymbol{F}\left(\omega_{1}\right)\times\boldsymbol{F}\left(\mathcal{\omega}_{2}\right)\right]\right\}\\
\times\left[\boldsymbol{F}\left(\omega_{3}\right)\cdot\boldsymbol{F}\left(\omega_{4}\right)\right]\left[\boldsymbol{F}\left(\omega_{5}\right)\cdot\boldsymbol{F}\left(\omega_{6}\right)\right]
\end{multline}
where we grouped the frequency arguments of $h^{(7)}$ with curly and squared brackets to improve readability. In this case there are new symmetries, we can exchange $\omega_{3}$
with $\omega_{5}$ and $\omega_{4}$ with $\omega_{6}$ simultaneously,
or $\omega_{3}$ with $\omega_{6}$ and $\omega_{4}$ with $\omega_{5}$
simultaneously. Again, the first step is to make sure that the triple
product is non-zero, which yields the same four triple products we
got for $h^{\left(5\right)}$ (see Methods). But this time, if we choose $\omega_{0}=-2\omega$,
$\omega_{1}=-\omega$, and $\omega_{2}=\omega$, instead of one we get five
different options that satisfy Eq. \eqref{eq:energy_conservation} for the rest of the frequencies :
\begin{eqnarray}
h_{a_{1}}^{\left(7\right)} & = & h^{\left(7\right)}\left(\{-2\omega,-\omega,\omega\},[\omega,\omega],[-\omega,\omega]\right), \label{eq:h7_a}\\
h_{a_{2}}^{\left(7\right)} & = & h^{\left(7\right)}\left(\{-2\omega,-\omega,\omega\},[-2\omega,2\omega],[\omega,\omega]\right), \\
h_{a_{3}}^{\left(7\right)} & = & h^{\left(7\right)}\left(\{-2\omega,-\omega,\omega\},[-\omega,-\omega],[2\omega,2\omega]\right), \\
h_{a_{4}}^{\left(7\right)} & = & h^{\left(7\right)}\left(\{-2\omega,-\omega,\omega\},[-2\omega,\omega],[\omega,2\omega]\right), \\
h_{a_{5}}^{\left(7\right)} & = & h^{\left(7\right)}\left(\{-2\omega,-\omega,\omega\},[-\omega,2\omega],[-\omega,2\omega]\right). 
\end{eqnarray}
Since the triple product can be written in three additional different
forms (see analogous discussion for $h^{\left(5\right)}$ in Methods) there are
a total of 20 different possible versions of $h^{\left(7\right)}$:
$h_{a_{i}}^{\left(7\right)}$, $h_{b_{i}}^{\left(7\right)}$, $h_{c_{i}}^{\left(7\right)}$,
and $h_{d_{i}}^{\left(7\right)}$, with $1\leq i\leq5$. If we now
consider the interaction with matter and limit the number of $2\omega$
photons to a maximum of one, i.e. we consider the $2\omega$ component
to be much weaker than the $\omega$ component $|F_z|\ll|F_x|$, we are left only with
$h_{a_{1}}^{\left(7\right)}$, $h_{b_{1}}^{\left(7\right)}$, $h_{c_{1}}^{\left(7\right)}$,
and $h_{d_{1}}^{\left(7\right)}$. These four options are related
to each other the same way that $h_{a}^{\left(5\right)}$, $h_{b}^{\left(5\right)}$,
$h_{c}^{\left(5\right)}$ and $h_{d}^{\left(5\right)}$ were related
to each other [see Eqs. \eqref{eq:ha}-\eqref{eq:hd_hc} and the corresponding discussion in Methods], and
therefore we arrive to the same conclusion as for $h^{\left(5\right)}$:
up to a sign and complex conjugation, there is a unique expression
for $h^{\left(7\right)}$ given by [see Eqs \eqref{eq:field_h5} in Methods]
\begin{equation}
h_{a_{1}}^{\left(7\right)}=2 i F_{x}F_{y}F_{z}\left(F_{x}^{2}-F_{y}^{2}\right)\left(F_{x}^{2}+F_{y}^{2}\right) e^{ 2i\left(\delta_{2\omega}-\delta_{\omega}\right)}.
\end{equation}
The next order chiral measure reads as
\begin{multline}
h^{\left(9\right)}\left(\left\{ \omega_{0},\omega_{1},\omega_{2}\right\} ,\left[\omega_{3},\omega_{4}\right],\left[\omega_{5},\omega_{6}\right],\left[\omega_{7},\omega_{8}\right]\right)=\left\{ \boldsymbol{F}\left(\omega_{0}\right)\cdot\left[\boldsymbol{F}\left(\mathcal{\omega}_{1}\right)\times\boldsymbol{F}\left(\omega_{2}\right)\right]\right\} \\ \times \left[\boldsymbol{F}\left(\omega_{3}\right)\cdot\boldsymbol{F}\left(\omega_{4}\right)\right]\left[\boldsymbol{F}\left(\omega_{5}\right)\cdot\boldsymbol{F}\left(\omega_{6}\right)\right]\left[\boldsymbol{F}\left(\omega_{7}\right)\cdot\boldsymbol{F}\left(\omega_{8}\right)\right]
\end{multline}
As for the seventh-order case, we obtain some new (trivial) symmetries derived from the commutativity of scalar products.  If we allow only a single $2\omega$ photon and choose $\omega_{0}=-2\omega$,
$\omega_{1}=-\omega$, and $\omega_{2}=\omega$ then the possible options for $h^{(9)}$ are
\begin{eqnarray}
h_{a_{1}}^{\left(9\right)} & = & h^{\left(9\right)}\left(\left\{ -2\omega,-\omega,\omega\right\} ,\left[\omega,\omega\right],\left[-\omega,\omega\right],\left[-\omega,\omega\right]\right),\\
h_{a_{2}}^{\left(9\right)} & = & h^{\left(9\right)}\left(\left\{ -2\omega,-\omega,\omega\right\} ,\left[\omega,\omega\right],\left[-\omega,-\omega\right],\left[\omega,\omega\right]\right).
\end{eqnarray}
Like before, since the triple product can be written in three additional
different forms there are a total of 8 different possible versions
of $h^{\left(9\right)}$ that involve a single $2\omega$ photon (absorbed
or emitted): $h_{a_{i}}^{\left(9\right)}$, $h_{b_{i}}^{\left(9\right)}$,
$h_{c_{i}}^{\left(9\right)}$, and $h_{d_{i}}^{\left(9\right)}$,
with $i=1,2$; related to each other as in the case of $h^{\left(5\right)}$
[see Eqs. \eqref{eq:ha}-\eqref{eq:hd_hc} and the corresponding discussion in Methods]. The explicit expressions for $h_{a_{1}}^{\left(9\right)}$
and $h_{a_{2}}^{\left(9\right)}$ read as [see Eq. \eqref{eq:field_h5} in Methods]
\begin{eqnarray}
h_{a_{1}}^{\left(9\right)} & = & 2 i F_{x}F_{y}F_{z}\left(F_{x}^{2}-F_{y}^{2}\right)\left(F_{x}^{2}+F_{y}^{2}\right)^{2} e^{ 2i\left(\delta_{2\omega}-\delta_{\omega}\right)}\\
h_{a_{2}}^{\left(9\right)} & = & 2 i F_{x}F_{y}F_{z}\left(F_{x}^{2}-F_{y}^{2}\right)^{3} e^{ 2i\left(\delta_{2\omega}-\delta_{\omega}\right)}
\end{eqnarray}
If we impose small ellipticity $\left|\epsilon\right|\equiv\left|F_{y}/F_{x}\right|\ll1$,
then to first order in $\epsilon$ we get $F_{x}^{2}\pm F_{y}^{2}\approx F_{x}^{2}$
and therefore $h_{a_{1}}^{\left(9\right)}=h_{a_{2}}^{\left(9\right)}$,
which leaves a unique expression for $h^{\left(9\right)}$ up to a
sign and complex conjugation:
\begin{equation}
h_{a_{1}}^{\left(9\right)}=2 i F_{x}^{7}F_{y}F_{z} e^{ 2i\left(\delta_{2\omega}-\delta_{\omega}\right)}.
\end{equation}
Higher-order measures will follow the same pattern and
will not introduce any new feature provided we enforce the restrictions
$\left|F_{y}\right|,\left|F_{z}\right|\ll\left|F_{x}\right|$, which is satisfied by the field employed in our work to demonstrate the highest possible degree of control over enantio-sensitive light matter interactions (see Figs. 3 and 4 of the main text).

\section {Another example of locally chiral field: counter-rotating bi-elliptical field}
\begin{figure}
\centering
\includegraphics[width=0.8\linewidth, keepaspectratio=true]{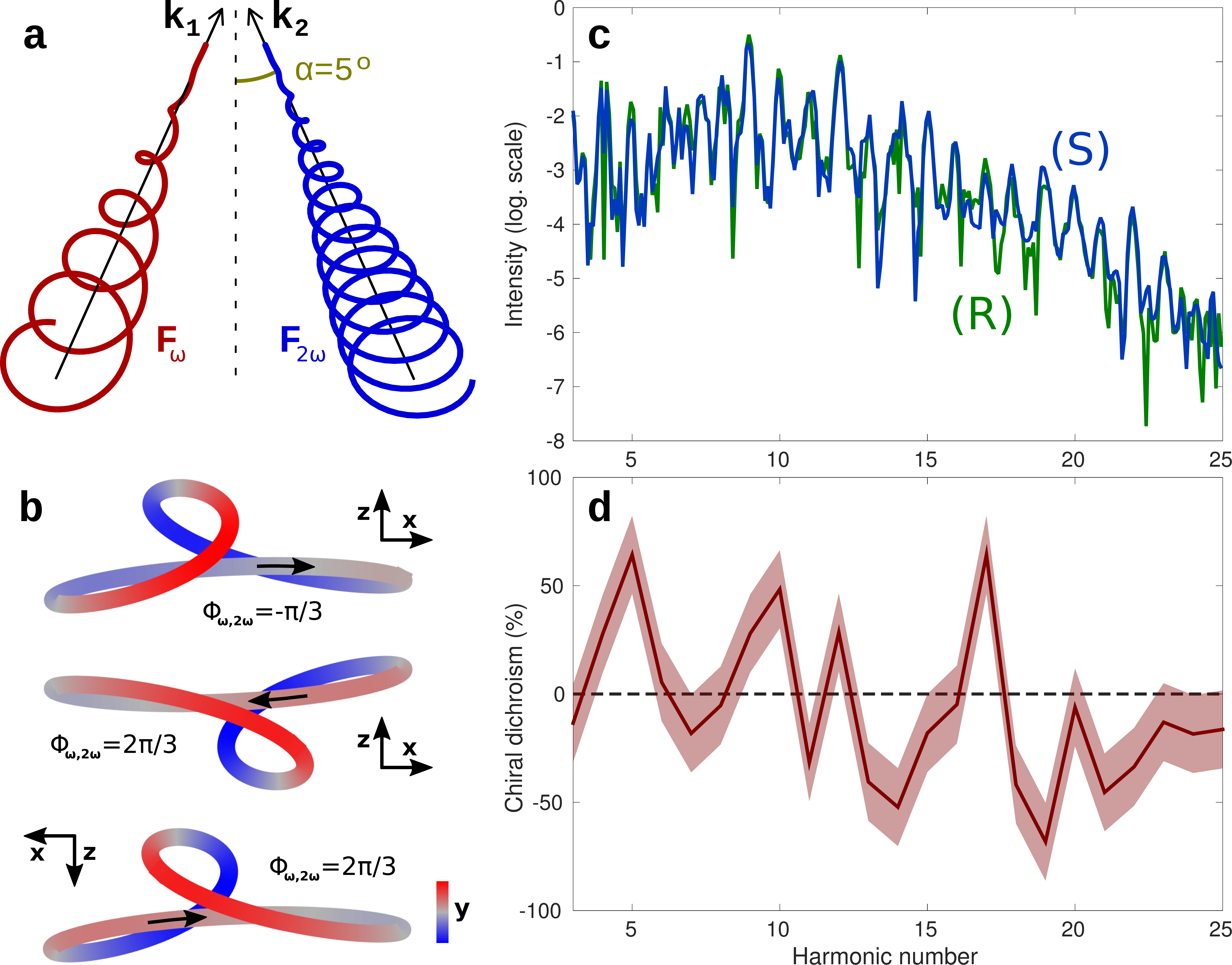}
\caption{Enantiosensitive response from a randomly oriented CBrClFH molecule.
a) Setup using counter-rotating elliptically polarized $\omega$ and $2\omega$ fields, generating the locally chiral field shown in panel (b) for $\phi_{\omega,2\omega}=-\pi/3$ (upper image) and $2\pi/3$  (central and bottom images).
The bottom image has been rotated $180^{\circ}$ around the $y$ axis to show that the fields for $\phi_{\omega,2\omega}=-\pi/3, 2\pi/3$ are mirror images.
c) $z$-polarized harmonic intensity from opposite enantiomers for $\lambda=1500$ nm, $I_{max}=1.2\times10^{13}$ W$/$cm$^2$, 
$\phi_{\omega,2\omega}=2\pi/3$, $\varepsilon_1=0.4$, $\varepsilon_2=0.3$, $\alpha=5^\circ$,
and a trapezoidal envelope with 4 cycle turn-on/off and 3 cycle flat-top.
d) Chiral dichroism in the harmonic intensity.
An error of $\pm 18.4\%$ is estimated based on convergence with the number of orientations.
}
\label{fig_bielliptical}
\end{figure}

The simple field shown in Fig. 1 a of the main text is only one example of a locally chiral field.
Fig. \ref{fig_bielliptical} a,b  presents an alternative locally chiral field that has different properties.
This field results from combining two counter-rotating elliptically polarized drivers with frequencies $\omega$ and $2\omega$ that propagate in different directions, with their polarization planes creating a small angle $\alpha$ (see Fig. \ref{fig_bielliptical}a).
The electric fields can be written at the focus as\cite{book_Boyd}
\begin{align}
\mathbf{F}_1(\mathbf{r},t) &= \frac{1}{2}F_{1,0} \hspace{0.2em} e^{-\rho_1^2/\tilde{\omega}^2} e^{i(\mathbf{k}_1\cdot\mathbf{r}- \omega t- \delta_{\omega} )} \hspace{0.2em} (\hat{\mathbf{e}}_1 + i\varepsilon_1 \hspace{0.2em} \hat{\mathbf{z}} ) + c.c. \\
\mathbf{F}_2(\mathbf{r},t) &= \frac{1}{2}F_{2,0} \hspace{0.2em} e^{-\rho_2^2/\tilde{\omega}^2} e^{i(\mathbf{k}_2\cdot\mathbf{r}-2\omega t-2\delta_{2\omega})} \hspace{0.2em} (\hat{\mathbf{e}}_2 - i\varepsilon_2 \hspace{0.2em} \hat{\mathbf{z}} ) + c.c. 
\end{align}
where 
$F_{n,0}$ is the electric field amplitude,
$\varepsilon_n$ is the ellipticity,
$\rho_n$ is the radial distance to beams' axis, 
$\tilde{\omega}$ is the waist radius,
the propagation vectors are defined as
$\mathbf{k}_1 =   k\sin{(\alpha)}\hat{\mathbf{x}} +  k\cos{(\alpha)}\hat{\mathbf{y}}$
and
$\mathbf{k}_2 = -2k\sin{(\alpha)}\hat{\mathbf{x}} + 2k\cos{(\alpha)}\hat{\mathbf{y}}$,
where $k=\frac{2\pi}{\lambda}$, $\lambda$ being the fundamental wavelength,
and the polarization vectors are given by
$\hat{\mathbf{e}}_1 = \cos{(\alpha)}\hat{\mathbf{x}} - \sin{(\alpha)}\hat{\mathbf{y}}$
and
$\hat{\mathbf{e}}_2 = \cos{(\alpha)}\hat{\mathbf{x}} + \sin{(\alpha)}\hat{\mathbf{y}}$.
The locally chiral field resulting from combining both beams can be written as
\begin{equation}\label{eq_field_bielliptical}
\mathbf{F}(\mathbf{r},t) = F_0 \hspace{0.2em} e^{-\rho_1^2/\tilde{\omega}^2} [ \mathbf{f}_1(\mathbf{r},t) + \mathbf{f}_2(\mathbf{r},t) ]
\end{equation}
where we have assumed $F_{0}=F_{1,0}=F_{2,0}$ and that $\rho=\rho_1\simeq\rho_2$ at the focus; $\mathbf{f}_1$ and $\mathbf{f}_2$ are
\begin{align}
\mathbf{f}_1(\mathbf{r},t) &= \frac{1}{2} e^{ i(\Phi_1 - \omega t)} \hspace{0.2em} (\hat{\mathbf{e}}_1 + i\varepsilon_1 \hspace{0.2em} \hat{\mathbf{z}} ) + c.c. \\
\mathbf{f}_2(\mathbf{r},t) &= \frac{1}{2} e^{2i(\Phi_2 - \omega t)} \hspace{0.2em} (\hat{\mathbf{e}}_2 - i\varepsilon_2 \hspace{0.2em} \hat{\mathbf{z}} ) + c.c. 
\end{align}
with
\begin{align}
\Phi_1 &=  k\sin{(\alpha)}x + k\cos{(\alpha)}y - \delta_{\omega} \\
\Phi_2 &= -k\sin{(\alpha)}x + k\cos{(\alpha)}y - \delta_{2\omega}
\end{align}
The handedness of this locally chiral field depends on the relative phase between the two colours
\begin{equation}\label{eq_delta_phi}
\Delta\Phi = \Phi_2 - \Phi_1 = -2k\sin{(\alpha)}x + \delta_{\omega} - \delta_{2\omega}
\end{equation}
Note that this relative phase does not depend on the direction of light propagation $y$ because $\frac{\mathbf{k}_{1}\cdot\mathbf{y}}{\omega_1} = \frac{\mathbf{k}_{2}\cdot\mathbf{y}}{\omega_2}$,
but it depends on the transversal coordinate $x$, as $\frac{\mathbf{k}_{1}\cdot\mathbf{x}}{\omega_1} \neq \frac{\mathbf{k}_{2}\cdot\mathbf{x}}{\omega_2}$.
Thus, the handedness of the locally chiral field changes along the $x$ direction and is not maintained globally in space.
Fig. \ref{fig_bielliptical}b of the main text shows that changing the phase shift $\Delta\Phi$ by $\pm\pi/2$ rad transforms the locally chiral field into its mirror image.
This means that the field has opposite handedness at positions $(x,y,z)$ and $(x+\Delta x,y',z')$, with $\Delta x = \frac{\lambda}{8\sin{(\alpha)}}$.

In order to illustrate that this locally chiral field can drive enantiosensitive response in chiral media, we have performed calculations for randomly oriented bromochlorofluoromethane molecules using Time Dependent Density Functional Theory, implemented in Octopus \cite{Marques2003,Andrade2015PCCP,Castro2006}.
We employed the Perdew-Burke-Ernzerhof exchange-correlation functional\cite{Perdew1996} of the generalized gradient approximation and pseudopotentials\cite{Schlipf2015} for the Br, Cl, F, and C atoms.

Fig. \ref{fig_bielliptical}c shows the single-molecule high harmonic response of enantiopure, randomly oriented media of left and right handed bromochlorofluoromethane molecules.
To demonstrate enantiosensitivity in odd harmonic frequencies, we have applied a polarization filter and show the intensity of the $z$-polarized radiation.
The calculated single-molecule high harmonic response shows extremely high degree of chiral dischroism in even and odd harmonic orders, reaching 60$\%$. The error bars at the level of 20$\%$ are associated with limited number of molecular orientations used for averaging over orientations.
The extreme computational cost of these calculations  makes optimization of enantio-sensitive response prohibitively expensive.  
In contrast to the field in Fig. \ref{fig_scheme}, this field is not globally chiral, as its handedness periodically alternates in space, see below.



\subsection{Analysis of local and global handedness of counter-rotating bi-elliptical field}
Here we apply the chiral correlation function $h^{(5)}$ [see Eq. \eqref{eq:h5_methods}] to illustrate the properties of the locally chiral field in Eq. \eqref{eq_field_bielliptical} from this perspective.
The analysis of the field correlation function $h^{(5)}$ shows that this locally chiral field carries different elements of chirality, which manifest itself in two types of correlation functions.
[see Eq. \eqref{eq:energy_conservation}]
\begin{eqnarray}
    h_{a}^{\left(5\right)}&=&h^{\left(5\right)}\left(-2\omega,-\omega,\omega,\omega,\omega\right)\label{eq:h5_a_haifa}\\
    h_{b}^{\left(5\right)}&=&h^{\left(5\right)}\left(-\omega,-2\omega,2\omega,2\omega,-\omega\right)\label{eq:h5_b_haifa}
\end{eqnarray}
Other options are related to either of these two by a change of sign, by complex conjugation, or by both as discussed in Methods. Note that the second option was not available for the field discussed in the main text because this option requires the $2\omega$ field to be elliptically polarized. Furthermore, it requires more than one $2\omega$ photon. Replacing Eq. \eqref{eq_field_bielliptical} in Eqs. \eqref{eq:h5_a_haifa} and \eqref{eq:h5_b_haifa} yields
\begin{equation}
h_{a}^{\left(5\right)}=2i\left(\frac{F_{0}}{2}\right)^{5}\varepsilon_{1}\left(1-\varepsilon_{1}^{2}\right)\sin\left(2\alpha\right) e^{i\left[4kx\sin\alpha+2\left(\delta_{2\omega}-\delta_{\omega}\right)\right]},
\label{h5a_haifa}
\end{equation}
\begin{equation}
h_{b}^{\left(5\right)}=-2i\left(\frac{F_{0}}{2}\right)^{5}\varepsilon_{2}\sin\left(2\alpha\right)\left[\cos\left(2\alpha\right)+\varepsilon_{1}\varepsilon_{2}\right]e^{i\left[-4kx\sin\alpha-2\left(\delta_{2\omega}-\delta_{\omega}\right)\right]}.
\label{h5b_haifa}
\end{equation}
Interestingly, $h^{(5)}_{a}$ is independent of the ellipticity of the $2\omega$ beam. We can also see from these expressions that we can obtain a locally chiral field with a unique $h^{(5)}$ by setting either of the two ellipticities to zero. That is, the fields with either $\varepsilon_{1}$ or $\varepsilon_{2}$ equal to zero are also locally chiral. Two different versions of $h^{(5)}$ mean that the field displays chirality at two different levels, i.e. like a helix made of a tighter helix. Finally, from these expressions it is clear that $h_{a}^{\left(5\right)}$ and $h_{b}^{\left(5\right)}$ oscillate as a function of $x$ with a frequency $4k\sin\alpha$, in agreement with the reasoning in the previous section. The global handedness, which obtains as a space-averaged value of correlation functions (Eq. \eqref{h5a_haifa}, \eqref{h5b_haifa}) is zero in case of this field.

\section{Control over chiral light matter interaction in the strong field regime}

\begin{figure}[H]
\centering
\includegraphics[width=\linewidth, keepaspectratio=true]{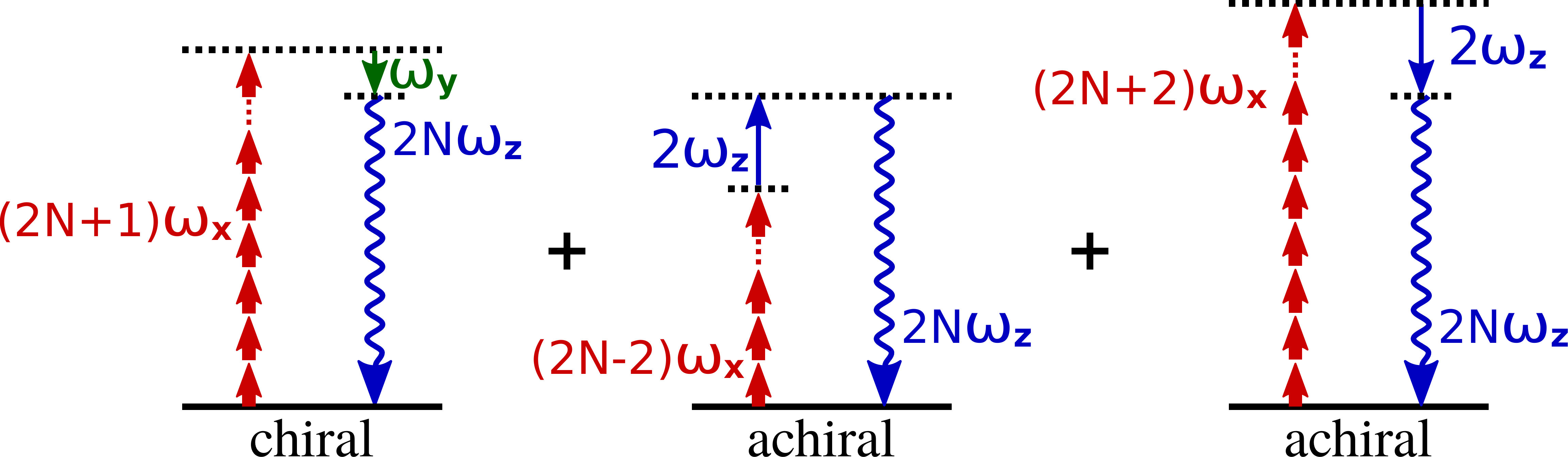}
\caption{Control over enantiosensitive high harmonic generation with locally chiral fields.
Interference of chiral (left diagram) and achiral (central and right diagrams) pathways in even high harmonic generation.}
\label{fig_SF_diagrams}
\end{figure}

Here we describe the mechanism of control over enantionsensitive high harmonic generation.
Even harmonic generation driven by the locally chiral field employed in the main text ( 1, 2, 3 and 4) results from the interference between the achiral and chiral pathways depicted in Fig. \ref{fig_SF_diagrams}.
The chiral pathway (left diagram) describes the enantiosensitive response driven by the elliptically polarized $\omega$ field in the direction orthogonal to the plane of polarization, which is not affected by the presence of the second harmonic field.

There are two possible achiral pathways giving rise to even harmonic generation.
One of them (central diagram in Fig. \ref{fig_SF_diagrams}) involves the absorption of $2N-2$ photons of $\omega$ frequency and $x$ polarization, absorption of a photon of $2\omega$ frequency and $z$ polarization and emission of a $2N$ harmonic with $z$ polarization.
The polarization associated with this process can be written as
\begin{equation}
\mathbf{P}_{2N}^{\uparrow} = P_{2N-2} \hspace{0.2em} \chi^{(1)}_{\uparrow} \hspace{0.2em} \mathbf{F}(2\omega) = a_{\uparrow} e^{i\delta} \hat{\mathbf{z}}
\end{equation}
where $P_{2N-2}$ is a scalar describing the absorption of $2N-2$ photons with $x$ polarization, that depends on the properties of the molecule and of the strong field component,
$\chi^{(1)}_{\uparrow}$ is the first order susceptibility of the system dressed by the $\omega$ field describing the absorption of a $2\omega$ photon with $z$ polarization,
and $a_{\uparrow} = P_{2N-2} \hspace{0.2em} \chi^{(1)}_{\uparrow} \hspace{0.2em} |\mathbf{F}(2\omega)\cdot\hat{\mathbf{z}}|$.
The relative phase between the two colours $\delta$ fully controls the phase of $\mathbf{P}_{2N}^{\uparrow}$.

Let us consider now the alternative achiral pathway (right diagram in Fig. \ref{fig_SF_diagrams}) involving the absorption of $2N+2$ $x$-polarized photons of $\omega$ frequency, emission of a $z$-polarized photon of $2\omega$ frequency and emission of a $z$-polarized $2N$ harmonic of the fundamental frequency.
The polarization term associated with this pathway is
\begin{equation}
\mathbf{P}_{2N}^{\downarrow} = P_{2N+2} \hspace{0.2em} \chi^{(1)}_{\downarrow} \hspace{0.2em} \mathbf{F}^*(2\omega) = a_{\downarrow} e^{-i\delta} \hat{\mathbf{z}}
\end{equation}
where $a_{\downarrow} = P_{2N+2} \hspace{0.2em} \chi^{(1)}_{\downarrow} \hspace{0.2em} |\mathbf{F}^*(2\omega)\cdot\hat{\mathbf{z}}|$.
The arrows $\uparrow$ and $\downarrow$ indicate whether the $2\omega$ photon is absorbed or emitted.

The total achiral contribution to polarization at $2N\omega$ frequency is given by $\mathbf{P}_{2N}^{\uparrow\downarrow} = \mathbf{P}_{2N}^{\uparrow} + \mathbf{P}_{2N}^{\downarrow}$.
If one of the two pathways is dominant, then $\delta$ fully controls the phase of $\mathbf{P}_{2N}^{\uparrow\downarrow}$.
If both pathways are equally intense, i.e. $a_{\uparrow}=a_0e^{i\phi_{\uparrow}}$ and $a_{\downarrow}=a_0e^{i\phi_{\downarrow}}$, then we have
\begin{equation}\label{eq_Pupdown}
\mathbf{P}_{2N}^{\uparrow\downarrow} = a_0 \Big( e^{i\phi_{\uparrow}} e^{i\delta} + e^{i\phi_{\downarrow}} e^{-i\delta} \Big) \hat{\mathbf{z}} = 2 a_0 e^{\phi_{+}} \cos{(\phi_{-}-\delta)} \hat{\mathbf{z}}
\end{equation}
where $\phi_{\pm}=\frac{\phi_{\downarrow}\pm\phi_{\uparrow}}{2}$.
Here we control the amplitude and the sign of the achiral contribution in full range.
The phase control is associated with the dependence of the phase of the recombination matrix element on the direction of electron approach (in the molecular frame).
Further, once one includes changes in ionization and recombination times due to the presence of the $2\omega$ field, one finds $\delta-$dependent corrections to the Volkov phase and thus the phase of a given harmonic.


\section*{Acknowledgements}
We thank Felipe Morales for stimulating discussions.
DA and OS acknowledge support from the DFG SPP 1840 ``Quantum Dynamics in Tailored Intense Fields'' and DFG grant SM 292/5-1; A.F.O. and OS acknowledge support from MEDEA.
The MEDEA project has received funding from the European Union's Horizon 2020 research and innovation programme under the Marie Sk\l{}odowska-Curie grant agreement No 641789.


\section*{Competing Interests}
The authors declare that they have no competing financial interests.
The data that support the plots within this paper and other findings of this study are available from the corresponding author upon reasonable request. 
Correspondence should be addressed to david.ayuso@mbi-berlin.de, mikhail.ivanov@mbi-berlin.de and olga.smirnova@mbi-berlin.de. 

\section*{References}
\bibliography{Bibliography}

\end{document}